\documentclass[11pt,a4paper]{article}
\pdfoutput=1
\usepackage{jheppub}
\usepackage{rotating}
\usepackage{array}
\usepackage{amsmath}
\usepackage[normalem]{ulem}
\usepackage{slashed}
\usepackage{booktabs}
\usepackage[pdftex,table]{xcolor}
\usepackage{units}

\newcommand{\be}{\begin{equation}}
\newcommand{\ee}{\end{equation}}
\newcommand{\bea}{\begin{eqnarray}}
\newcommand{\eea}{\end{eqnarray}}
\newcommand{\hd}{{\sf HDECAY}}
\newcommand{\ds}{{\sf DarkSUSY}}
\newcommand{\code}[1]{{\tt #1}}

\usepackage{multirow}

\preprint{TTK-21-48}

\title{Freezing-in a hot bath: 
resonances, medium effects and phase transitions}

\author[1,2]{Torsten Bringmann,}
\author[3,4]{Saniya Heeba,}
\author[3]{Felix Kahlhoefer}
\author[1]{and Kristian Vangsnes}

\affiliation[1]{Department of Physics, University of Oslo, Box 1048, N-0316 Oslo, Norway}
\affiliation[2]{AEC, Institute for Theoretical Physics, University of Bern, Sidlerstrasse 5, CH-3012 Bern, Switzerland}
\affiliation[3]{Institute for Theoretical Particle Physics and Cosmology (TTK), RWTH Aachen University, D-52056 Aachen, Germany}
\affiliation[4]{Department of Physics \& McGill Space Institute, McGill University, Montr\'eal, QC H3A 2T8, Canada}

\emailAdd{torsten.bringmann@fys.uio.no}
\emailAdd{saniya.heeba@mcgill.ca}
\emailAdd{kahlhoefer@physik.rwth-aachen.de}
\emailAdd{k.g.vangsnes@fys.uio.no}

\abstract{%
Relic density calculations of dark matter freezing out from the primordial plasma 
have reached a high level of sophistication, with several numerical tools readily available 
that match the observationally required accuracy. 
 Dark matter production via the freeze-in mechanism, on the other hand, 
 is sensitive to much higher temperatures than in the freeze-out case, implying both
 technical and computational difficulties when aiming for the same level of precision. 
 We revisit the formulation of freeze-in production in a way that facilitates the inclusion
 of in-medium corrections like plasma effects and the spin statistics of relativistic quantum 
 gases, as well as the temperature dependence of dark matter production rates
 induced by the electroweak and strong phase transitions, and we discuss in detail
 the additional complications arising in the presence of $s$-channel resonances.
 We illustrate our approach in the context of Higgs portal models, and provide the 
most accurate calculation to date of the freeze-in abundance of Scalar Singlet dark matter. We 
explore in particular the case of small reheating temperatures, for which the couplings implied by the 
freeze-in mechanism may be testable at the LHC.
Together with this article we present a major update 6.3 of \ds\ with the added capability
of performing general freeze-in calculations, including all complications mentioned above.
}

\keywords{Mostly Weak Interactions: Beyond Standard Model; Astroparticles: Cosmology of Theories beyond the SM, Thermal Field Theory}

\begin{document}
\maketitle
\flushbottom

\section{Introduction}
\label{sec:introduction}

One of the great successes of modern cosmology is its ability to describe the universe at the largest 
scales in terms of the properties of the elementary particles it contains. Famously, cosmological 
observations tightly constrain the effective number of neutrino species in agreement with the 
predictions of the Standard Model (SM)~\cite{Planck:2018vyg}. A similar success story is anticipated 
in the case of dark matter, whose relic density in the universe is precisely measured although its 
fundamental properties are still unknown.

Indeed, for a given particle physics model of DM it is typically possible to calculate the predicted relic 
density with great precision. In many cases this number is given by the freeze-out mechanism, 
i.e.\ the decoupling of DM particles from chemical equilibrium. Various public numerical codes, 
such as \ds~\cite{Bringmann:2018lay}, {\sf micrOmegas}~\cite{Belanger:2018ccd} and 
{\sf MadDM}~\cite{Ambrogi:2018jqj} have been developed to automate the relic density calculation 
even in complicated scenarios, including for example co-annihilations, thresholds and resonances. 

At the same time, an increasing amount of attention has been paid to DM models where one cannot 
simply assume chemical equilibrium between all particle species at early times. The most well-known 
example is the freeze-in mechanism~\cite{Hall:2009bx,Chu:2011be,Chu:2013jja}, in which 
interaction rates are so small compared to the Hubble expansion rate that the DM density evolves 
towards an equilibrium distribution without ever reaching it. Motivation for such models partially 
stems from 
the non-observation of convincing DM signals, which is difficult to reconcile with the generic 
predictions of the freeze-out mechanism and points towards feebly-interacting 
particles (FIMPs)~\cite{Bernal:2017kxu}.

What makes relic density calculations in the freeze-in  framework particularly challenging is that the 
final DM density is typically sensitive to a wide range of temperatures in the early universe, including 
in particular temperatures well above the DM mass~\cite{Lebedev:2019ton,Biondini:2020ric}. In fact, 
in many models of freeze-in there may even be a dependence on initial conditions, such as the 
details of reheating~\cite{Hardy:2017wkr}. At such high temperatures a number of new effects 
become relevant, specifically in-medium corrections like plasma 
effects~\cite{Dvorkin:2019zdi,Heeba:2019jho,Hambye:2019dwd,Darme:2019wpd} and the spin 
statistics of relativistic quantum gases~\cite{Belanger:2018ccd,Bandyopadhyay:2020ufc}, as well as 
phase transitions, which can fundamentally change the relevant degrees of freedom of the theory 
under consideration~\cite{Baker:2017zwx,Heeba:2018wtf}. Including all of these effects in order to 
obtain precise predictions for the DM relic density is technically and computationally difficult.

The present work addresses these issues by formulating the freeze-in formalism in a way that can be 
straight-forwardly implemented in numerical codes developed to study the freeze-out mechanism. 
Key to this approach is to rewrite the DM production rate in terms of the DM annihilation rate, which 
is the central quantity for the freeze-out mechanism. We show how to consistently include in-medium 
effects in this reformulation and discuss in detail the complications arising from $s$-channel 
resonances. All of these effects have been implemented in the most recent version 6.3~of \ds, 
which will be released together with this work.
This release makes \ds\ the second publicly available code (after micrOmegas) to provide freeze-in 
routines for general DM models, and the first one to take into account all relevant in-medium effects,
including those induced by the SM phase transitions.

We illustrate our approach in the context of Higgs portal models (see ref.~\cite{Lebedev:2021xey} 
for a recent review), which in spite of their simplicity turn out to require highly complex freeze-in 
calculations. The reason is that the relic density depends directly on the off-shell decay width of the 
SM Higgs boson at finite temperatures. At centre-of-mass energies well above the Higgs boson 
mass, great care is required when including higher-order corrections to avoid an unphysical growth 
of the cross section. At centre-of-mass energies well below the Higgs boson mass, on the other 
hand, it is crucial to consider the transition from free quarks and gluons in the final state to hadronic 
bound states.

We apply all these findings to the well-studied case of scalar singlet DM~\cite{Silveira:1985rk,McDonald:1993ex,Burgess:2000yq}, which has been studied in detail both for 
freeze-out~\cite{Cline:2013gha,GAMBIT:2017gge,Athron:2018ipf} and 
freeze-in~\cite{Yaguna:2011qn} production, and provide the most accurate calculation to date of the relic 
abundance of these particles via the freeze-in mechanism. In agreement with previous 
results~\cite{Belanger:2018ccd}, we find relatively small corrections in the case that DM particles are 
dominantly produced in Higgs decays, but point out that corrections can be more significant for larger 
masses. Moreover, we consider for the first time the case that the reheating temperature is smaller 
than the Higgs boson mass, such that DM production proceeds dominantly via an effective 
dimension-5 operator. In this case, much larger couplings are required to reproduce the observed 
DM relic abundance, which may be testable via precision measurements of the branching ratios of 
the observed SM-like Higgs boson.

The remainder of this work is structured as follows. In section~\ref{sec:freeze-in} we present our 
formulation of the freeze-in formalism in terms of DM annihilations. We take special care to include 
in-medium effects and discuss the appropriate treatment of $s$-channel resonances. In 
section~\ref{sec:finiteT} we then take a closer look at finite-temperature effects, with a particular 
focus on the temperature-dependent Higgs vacuum expectation value (vev) and phase transitions in 
the early universe. Section~\ref{sec:hdecay} is dedicated to a detailed discussion of the decays of 
off-shell Higgs bosons, both in the case where the CMS energy is much larger and much smaller 
than the Higgs boson rest mass, respectively. The importance of all of these aspects is then 
exemplified in section~\ref{sec:scalarsinglet}, where we consider the freeze-in production of scalar 
singlets for different assumptions on the reheating temperature. 
We conclude in section~\ref{sec:conclusions} with a summary of our main findings
and possible implications for future investigations.
In appendix \ref{app:ds} we describe the
implementation of the new freeze-in routines in \ds, and appendix \ref{app:technical}
complements section~\ref{sec:freeze-in} by providing further technical details.

\section{Freeze-in formalism}
\label{sec:freeze-in}
We start with a general description of the freeze-in process~\cite{Hall:2009bx}, 
i.e.~the thermal production of DM particles with interaction strengths too weak to
ever equilibrise with the heat bath. 
To keep the discussion in this section general, we allow for both, DM particles $\chi$ 
and heat bath (SM) particles $\psi$, to have arbitrary mass and spin.
We put special emphasis on the fact that the DM production from the heat bath through $2\to2$ processes, 
$\psi\psi\to\chi\chi$,
can equivalently be described in terms of the annihilation of a {\it would-be} thermal population
of DM particles, $\chi\chi\to\psi\psi$. As we demonstrate below, this formal equivalence not only
holds when assuming Maxwell-Boltzmann distributions -- as familiar from cold DM freeze-out 
scenarios~\cite{Gondolo:1990dk} --
but even when fully taking into account the effect of quantum statistics in the phase-space distributions
of the involved particles.

\subsection{Boltzmann equation}
The evolution of the number density $n_\chi$ of DM particles in the early universe
is governed by the Boltzmann equation
\be
  \label{boltzmann}
  \dot n_\chi +3 H n_\chi = C[f_\chi] \,,
\ee
where $\dot{ }\equiv\, d/dt$, $H=\dot a/a$ is the Hubble rate,  $f_\chi$ denotes the phase-space density of $\chi$ 
and $C[f_\chi]$ is the collision operator for all processes that do not conserve the number of $\chi$ particles. 
We will focus on the interactions between two DM particles with 4-momenta $(E, \mathbf{p})$ and 
$(\tilde{E},\mathbf{\tilde{p}})$ and two SM particles with 4-momenta $(\omega, \mathbf{k})$ and 
$(\tilde{\omega},\mathbf{\tilde{k}})$. In the cosmic rest frame the collision operator then takes the general 
form~\cite{Kolb:1990vq}
\bea
  \label{C_def}
  C[f_\chi]&=&\frac1{N_\psi}\int\!\!\frac{d^3 p}{(2\pi)^3 2E}\int\!\!\frac{d^3\tilde p}{(2\pi)^32\tilde E}\int\!\!\frac{d^3k}{(2\pi)^32\omega}\int\!\!\frac{d^3\tilde k}{(2\pi)^32\tilde \omega}\,(2\pi)^4\delta^{(4)}(\tilde p+p-\tilde k-k)\\
&&\times\left[
\left|\mathcal{M}\right|^2_{\chi\chi\leftarrow \psi\psi}f_\psi(\omega)f_\psi(\tilde \omega)
\bar f_\chi(E) \bar f_\chi(\tilde E) 
-\left|\mathcal{M}\right|^2_{\chi\chi\rightarrow \psi\psi}f_\chi(E)f_\chi(\tilde E) \bar f_\psi(\omega) \bar f_\psi(\tilde \omega)
\right]\,,\nonumber
\eea
where $f_{\chi,\psi}$ denote the phase-space distribution functions of $\chi$ and $\psi$, and the 
factors
\be
\bar f_i\equiv1-\varepsilon_i f_i
\ee
 reflect in-medium effect due to quantum statistics, i.e.~Pauli blocking for fermions 
($\varepsilon_{\chi,\psi}=+1$) and Bose enhancement for Bosons ($\varepsilon_{\chi,\psi}=-1$) in the 
final state. We further introduced an explicit factor of $N_\psi=2$ $(1)$ for self-conjugate 
(not self-conjugate) SM particles $\psi$, i.e.~we use a convention where each of the phase-space 
integrals is always understood to be performed over all possible momentum configurations.
The scattering amplitude $\mathcal{M}$ is squared and then summed over both initial and final state 
degrees of freedom; assuming $CP$ invariance furthermore allows us to introduce 
${\left|\mathcal{M}\right|}^2\equiv\left|\mathcal{M}\right|^2_{\chi\chi\leftarrow \psi\psi}=\left|\mathcal{M}\right|^2_{\chi\chi\rightarrow \psi\psi}$.
We emphasise that all 4-momenta in the expression for $C[f_\chi]$ have to be evaluated in 
the cosmic rest frame, as it is only in this frame that the distribution functions are guaranteed to have 
no angular dependence. The phase-space distribution of the heat bath particles, in particular, is 
given by the usual $f_\psi(\omega)=1/\left[\exp(\omega/T)+ \varepsilon_\psi\right]$, 
where $T$ is the photon temperature.

We will here exclusively be interested in the {\it freeze-in regime} of the above expression, characterised by
two independent requirements on the DM distribution:
\begin{enumerate}
 \item $f_\chi\ll1$: The DM abundance -- assumed to vanish initially -- remains so small that Pauli blocking, 
 or Bose enhancement, is irrelevant for the first term in eq.~(\ref{C_def}). 
  \item $f_\chi\ll g$: The DM abundance stays sub-thermal, implying that the effect of DM
  annihilations -- the second term in eq.~(\ref{C_def}) -- is negligible.
\end{enumerate}
In the freeze-in regime, the collision term thus becomes completely independent of the DM phase-space distribution.
For this reason, it is conventionally expressed in terms of the DM {\it production} cross section
$\sigma_{\psi\psi\to\chi\chi}$.

In this article we follow a different approach and express the collision term in the freeze-in regime
in terms of the DM {\it annihilation} cross section $\sigma_{\chi\chi\to \psi\psi}$ (see also 
refs.~\cite{Arcadi:2019oxh,Lebedev:2019ton}).
To do so, we first note that energy conservation implies
\be
\label{eq:thermo_identity}
f_\psi(\omega)f_\psi(\tilde\omega)=
f_\psi(\omega)f_\psi(\tilde\omega) e^{(\omega+\tilde\omega)/T} e^{-(E+\tilde E)/T}
= 
\bar f_\psi(\omega)\bar f_\psi(\tilde \omega)
f_\chi^{\rm MB}(E)\,f_\chi^{\rm MB}(\tilde E)\,,
\ee
where we have introduced 
\be
f_\chi^{\rm MB}(E)\equiv \exp(-E/T)\,.
\ee
The production term in eq.~(\ref{C_def}) thus takes the same form as the annihilation term
{\it would} take for a fiducial DM phase-space density following a Maxwell-Boltzmann distribution.
We stress that in arriving at this result we did not make any assumptions about the {\it actual} phase-space
distribution of the DM particles, other than $f_\chi\ll1$ (assumption 1 above). 
This implies that the
r.h.s.~of the Boltzmann equation for the number density can be written as
\be
\label{eq:coll_prod}
C[f_\chi] = \langle \sigma v\rangle_{\chi\chi\to\psi\psi} \left(n_\chi^{\rm MB}\right)^2,
\ee
where $n_\chi^{\rm MB}\equiv g_\chi(2\pi)^{-3}\int d^3p\,f_\chi^{\rm MB}=g_\chi m_\chi^2 T K_2(m_\chi/T)/(2\pi^2)$,
with $K_2$ a modified Bessel function of the second kind, and
\be
\label{eq:svav_def}
 \langle \sigma v\rangle_{\chi\chi\to\psi\psi} \equiv 
  \frac{g_\chi^2}{\left(n_\chi^{\rm MB}\right)^{2}} \int\!\!\frac{d^3 p}{(2\pi)^3}\int\!\!\frac{d^3\tilde p}{(2\pi)^3} f_\chi^{\rm MB}(E)
  f_\chi^{\rm MB}(\tilde E)\, v_{\rm M\o l} \sigma_{\chi\chi\to\psi\psi}(p,\tilde p)\,.
\ee
Here, $v_{\rm M\o l}\equiv F/(E \tilde E) \equiv \sqrt{(p\cdot\tilde p)^2-m_\chi^4}/(E \tilde E)$ is the M\o ller velocity and 
$\sigma_{\chi\chi\to\psi\psi}$ is the in-medium annihilation cross section in the cosmic rest frame,
i.e.~taking into account the effect of quantum statistics in the final state:
\be
\label{sigma_def}
\sigma_{\chi\chi\to\psi\psi} (p,\tilde p) = 
\frac{(2\pi)^4}{4 N_\psi F}
\int\!\!\frac{d^3k}{(2\pi)^32\omega}\int\!\!\frac{d^3\tilde k}{(2\pi)^32\tilde \omega}\,\delta^{(4)}(\tilde p+p-\tilde k-k)
\left|\overline{\mathcal{M}}\right|^2
\bar f_\psi(\omega) \bar f_\psi(\tilde \omega)\,,
\ee
where the spin-averaged amplitude squared is as usual denoted as 
$\left|\overline{\mathcal{M}}\right|^2\equiv \left|{\mathcal{M}}\right|^2/g_\chi^2$.

Let us briefly pause, and compare our result to the situation in the standard freeze-out 
scenario~\cite{Gondolo:1990dk} where,
formally, the DM production term is identical to that in eq.~(\ref{eq:coll_prod}). The physical difference is two-fold:
{\it i)} during the freeze-out of non-relativistic particles, $f_\chi^{\rm MB}$ describes the {\it actual} equilibrium
distribution, and {\it ii)} in-medium effects due to quantum statistics 
are irrelevant for the annihilation cross section; this is because energy 
conservation restricts the SM phase-space densities to their
high-energy tails, thus effectively implementing `$\varepsilon_\psi=0$' in eq.~(\ref{sigma_def}).
Still, as we will demonstrate below, the fact that eqs.~(\ref{eq:coll_prod}, \ref{eq:svav_def}) take the 
same form as in the freeze-out case is highly beneficial both from the point of view of the numerical 
implementation and when estimating higher-order corrections to the scattering cross sections.

\subsection{Relativistic collision operator for quantum gases}
Evaluating the phase-space integrals appearing in eq.~(\ref{eq:svav_def}) is most easily done
in the centre-of-mass (CMS) frame. This has the additional advantage that the final result will also depend on the
annihilation cross section in that frame (or any other frame boosted along the collision axis),
and thus on the standard frame in which cross sections are typically stated. 
From now on, $\sigma_{\chi\chi\to\psi\psi}$ will thus always refer to the CMS cross section;
in particular, we will drop the explicit dependence on $(p,\tilde p)$ to avoid confusion with the cross
section in the cosmic frame appearing in eq.~(\ref{eq:svav_def}).
Neglecting quantum statistics factors in eq.~(\ref{sigma_def}), 
then results in the often quoted expression for the thermally averaged annihilation cross section as 
derived by Gelmini and Gondolo~\cite{Gondolo:1990dk}:
\be
 \label{eq:svav_GG}
 \langle \sigma v\rangle^{\rm GG}
 =  \int_1^\infty\!\!\! d\tilde s\,
 \frac{4 x\sqrt{\tilde s}({\tilde s-1})\, K_1\!\left({2{\sqrt{\tilde s}} x}\right)}
 {{K_2}^2(x)}
  \sigma^{\epsilon_\psi\to0}_{\chi\chi\rightarrow \psi\psi}\,,
\ee
where $x\equiv m_\chi/T$ and $\tilde s \equiv s/(4m_\chi^2)$ are dimensionless parameters, $\sqrt{s}$ being the total CMS energy. 

The simplicity of the result obtained by Gelmini and Gondolo is a direct consequence of the fact
that $\sigma_{\chi\chi\rightarrow \psi\psi}^{\varepsilon_\psi\to0}$ is only a function of $s$; in particular, the 
phase-space integrals in eq.~(\ref{sigma_def}) do not introduce any frame-dependence in that case.
This changes radically when including the heat bath distribution functions, which take the simple form 
$f_\psi(\omega)=1/\left[\exp(\omega/T)+ \varepsilon_\psi\right]$ only in the cosmic rest frame.
In other words, in order to calculate the full cross section in the CMS frame, including quantum statistics, 
we need to know how 
the CMS frame relates to the cosmic rest frame. Due to the isotropy of space, it must be possible to state 
this relation in terms of a single boost parameter between the two frames, for which we will use 
the rapidity $\eta$:
\be
\sigma_{\chi\chi\to\psi\psi} =\sigma_{\chi\chi\to\psi\psi} (s,\eta)\,.
\ee

In deriving $\sigma_{\chi\chi\to\psi\psi} (s,\eta)$ and its thermal average in a closed form, in analogy 
to eq.~(\ref{eq:svav_GG}), we will heavily borrow from the treatment presented 
in refs.~\cite{Arcadi:2019oxh,Lebedev:2019ton}. In particular, we note that
 the phase-space distribution $f_\psi(\omega)$  becomes $f_\psi(u \cdot k)$ in a general frame,
where $u$ is the 4-velocity of the cosmic fluid, with $u \cdot k = \omega \cosh \eta + k^3 \sinh \eta$ 
in the CMS frame (and likewise for $f_\psi(\tilde{\omega})$).
Using further that in this frame we have $k^3=-\tilde k^3=\cos\theta\, |\mathbf{k}_{\rm CM}|$, we find that the 
plasma-frame dependent factors in eq.~(\ref{sigma_def}) are captured in the quantity
\bea
\label{eq:Gsimp}
G^{-1}(\gamma, s, \cos\theta)
&\equiv& \left[\bar f_\psi(u \cdot k )\bar f_\psi(u \cdot \tilde k )\right]^{-1}\\
&=&
1+\varepsilon_\psi^2e^{-2\sqrt{\tilde{s}} x \gamma}
-2\varepsilon_\psi e^{-\sqrt{\tilde{s}} x \gamma}
\cosh\left[
\cos\theta \left(\sqrt{\tilde{s}}-\frac{m_\psi^2}{m_\chi^2}\right)^\frac12 x\sqrt{\gamma^2-1} 
\right],\nonumber
\eea
where we have introduced the Lorentz factor $\gamma\equiv \cosh\eta$ for later computational ease;
we also kept an explicit factor of $\varepsilon_\psi^2$
such that formally setting `$\varepsilon_\psi=0$' in the above expression correctly reproduces $G=1$ 
(as expected in the absence of plasma effects due to quantum statistics).
The phase-space integrals in eq.~(\ref{sigma_def}) can then as usual~\cite{Peskin:1995ev} be reduced to one angular
integral, such that the final expression for the in-medium annihilation cross section in the CMS frame becomes
\be
\label{eq:sigfinal}
\sigma_{\chi\chi\to\psi\psi} (s,\gamma)=
\frac{N_\psi^{-1}}{8\pi s}\frac{|\mathbf{k}_{\rm CM}|}{\sqrt{s-4m_\chi^2}}
\int_{-1}^{1}\frac{d\cos\theta}{2}\left|\overline{\mathcal{M}}\right|^2\!(s,\cos\theta)\, G(\gamma, s, \cos\theta)\,.
\ee
It is further worth noting that  the phase-space integration of an arbitrary function $f_\psi(p,\tilde p)$ can be
rewritten as~\cite{Arcadi:2019oxh}
\be
\label{eq:oleg_trick}
\int\!\!\frac{d^3 p}{2 E}\int\!\!\frac{d^3\tilde p}{2\tilde E}\, f_\psi(p,\tilde p)=
m_\chi^4\int_1^\infty d\tilde s\, \sqrt{\tilde s(\tilde s-1)} \int_0^\infty d\eta\,\sinh^2\eta\,
\int d\Omega_{\bar p}d\Omega_{\bar k}\, \left.f_\psi(p,\tilde p)\right|_{k_0=0}\,,
\ee
after changing variables to $\bar p\equiv(p+\tilde p)/2$ and $\bar k\equiv(p-\tilde p)/2$,
with $\bar p^0\equiv ({\sqrt{s}}/{2})\cosh\eta$.
Here, $\Omega_{\bar p}$ $(\Omega_{\bar k})$ is the solid angle w.r.t.~$\bar p$ $(\bar k)$
and we note that $k_0=0$ implies $|\bar{\mathbf{k}}|=m_\chi\sqrt{\tilde s-1}$. 
This allows to rewrite the thermal average appearing in eq.~(\ref{eq:svav_def})
in the following, compact form:
\be
\label{eq:svav_def_full}
 \langle \sigma v\rangle_{\chi\chi\to\psi\psi}
 =\frac{8x^2}{{K_2}^2(x)}\int_1^\infty d\tilde s\,\tilde s\left(\tilde s -1\right) \int_1^\infty d\gamma\, \sqrt{\gamma^2-1}e^{-2\sqrt{\tilde{s}} x \gamma}
 \sigma_{\chi\chi\to\psi\psi} (s,\gamma)\,.
\ee

Eq.~(\ref{eq:svav_def_full}), along with  eqs.~(\ref{eq:Gsimp}, \ref{eq:sigfinal}), constitutes one of our central results.
We stress that it is valid under fairly general conditions, and applies independently of whether either of the $\chi$ or $\psi$
particles is self-conjugate or not. In fact, eqs.~(\ref{eq:sigfinal}, \ref{eq:svav_def_full}) take the 
same form also for DM annihilation processes where the two final-state (SM) particles have different masses; 
only the expression for $G(\gamma, s, \cos\theta)$ in eq.~(\ref{eq:Gsimp}) has to be replaced in that case,
and we state the corresponding generalised version in appendix \ref{app:technical}. In the same appendix, we 
also provide analytical results for the angular integration
in eq.~(\ref{eq:sigfinal}) for the case of amplitudes without dependence on the scattering angle, 
$\left|\overline{\mathcal{M}}\right|^2=\left|\overline{\mathcal{M}}\right|^2\!(s)$.
We finally note that without in-medium effects due to quantum statistics, i.e.~for $G\equiv1$, the 
integral over $\gamma$ in eq.~(\ref{eq:svav_def_full})
can be performed analytically, leading as expected to the familiar result stated in eq.~(\ref{eq:svav_GG}).

To summarise this Section so far, the Boltzmann equation governing the evolution of the $\chi$ number density 
in the freeze-in regime can always be written as
\be
  \label{eq:boltzmann}
  \dot n_\chi +3 H n_\chi = \langle \sigma v\rangle \left(n_\chi^{\rm MB}\right)^2\,,
\ee
where $\chi$ may be self-conjugate ($\chi=\bar\chi$) or not, and we introduced the total DM annihilation rate
as $\sigma \equiv\sum_{i,j}\sigma _{\chi\bar{\chi}\to\psi_i\psi_j}$, the sum being over all heat bath particles $\psi_i$.
Despite its appearance, this equation fully takes into account both relativistic kinematics and
the effect of quantum statistics. Writing it in this form, thus stressing the formal analogy
with the production term for freeze-out in the non-relativistic limit, is clearly advantageous from a numerical implementation 
point of view, c.f.~appendix \ref{app:ds}; as we will see in Section \ref{sec:hdecay}, furthermore,
it also allows a more sophisticated treatment of DM production from the heat bath
through an off-shell Higgs resonance (compared to what is easily achievable with the standard formulation).
It is also worth stressing that, in contrast to the freeze-out situation, 
the above Boltzmann equation for $n_\chi$ can be straight-forwardly solved by direct 
integration. This becomes more apparent when rewriting it as
\be
\frac{dY_\chi}{dx}=\frac{\left(n_\chi^{\rm MB}\right)^2}{x s \tilde H}\langle \sigma v\rangle\,,
\label{eq:dYdx}
\ee
where we have assumed entropy conservation and denoted the abundance of $\chi$ as $Y_\chi\equiv n_\chi/s$, 
with $s$ being the entropy density; we also introduced 
$\tilde H\equiv H/\left[1+ (1/3)d(\log g^s_{\rm eff})/d(\log T)\right]$ and the effective entropy 
degrees of freedom, $g^s_{\rm eff}$. Integrating this equation for $x\to\infty$ then gives the abundance of $\chi$ 
today, $Y^0_\chi$, which is related to the observed DM density as 
$\Omega_{\rm DM} h^2= 2.755\times10^{10}\left({m_\chi}/{100\,{\rm GeV}}\right) 
\left(2/N_\chi \right)
Y^0_\chi$,
with $N_\chi=2$ $(1)$ for self-conjugate (not self-conjugate) DM particles $\chi$.

\subsection{Dark matter production from decay}
\label{sec:FI_decays}

While our emphasis is on $2\to2$ processes, 
we note that freeze-in production of DM is also possible through 
the decay $A\to\chi\chi$ of some bosonic particle $A$. 
The general form of the collision operator for decay is given by
\bea
  \label{C_decay_def}
  C_{\rm dec}[f_\chi]&=&\frac1{N_\chi}\int\!\!\frac{d^3 p}{(2\pi)^3 2E}\int\!\!\frac{d^3\tilde p}{(2\pi)^32\tilde E}\int\!\!\frac{d^3k'}{(2\pi)^32\omega'}\,(2\pi)^4\delta^{(4)}(\tilde p+p-k')\nonumber\\
&&\times\Bigg[
\left|\mathcal{M}\right|^2_{A\rightarrow\chi\chi}f_\psi(\omega')
\bar f_\chi(E) \bar f_\chi(\tilde E) -\left|\mathcal{M}\right|^2_{A\leftarrow\chi\chi}f_\chi(E)f_\chi(\tilde E) \bar f_\psi(\omega')
\Bigg]\\
&=&\frac1{N_\chi}\int\!\!\frac{d^3 p}{(2\pi)^3 2E}\int\!\!\frac{d^3\tilde p}{(2\pi)^32\tilde E}\int\!\!\frac{d^3k'}{(2\pi)^32\omega'}\,(2\pi)^4\delta^{(4)}(\tilde p+p-k') \left|\mathcal{M}\right|^2_{A\rightarrow\chi\chi}f_A(\omega')\,,\nonumber\\
\eea
where $(\omega', \mathbf{k}')$ denotes the 4-momentum of $A$ and the second step
is essentially a definition of the freeze-in regime, in analogy to what we did for $2\to2$
processes.
A particularly common application of this expression is the situation where $A$ is in thermal 
equilibrium with the heat bath, i.e.~where $f_A(\omega')$ is given by a Bose-Einstein distribution. 
Using a similar argument as in eq.~(\ref{eq:thermo_identity}), we can then  rewrite 
the collision term in a form that appears to describe an inverse decay process $\chi\chi\to A$ from 
a {\it fiducial} Maxwell-Boltzmann distribution of DM particles:
\be
\label{C_inverse_decay}
  C_{\rm dec}[f_\chi]=\frac1{N_\chi}\int\!\!\frac{d^3 p}{(2\pi)^3 2E}\int\!\!\frac{d^3\tilde p}{(2\pi)^32\tilde E}
  \left|\mathcal{M}\right|^2_{A\rightarrow\chi\chi} f_\chi^{\rm MB}(E) f_\chi^{\rm MB}(\tilde E)
  \frac{\pi}{\omega'}\delta(\omega'-E-\tilde E)\,  \bar f_A(\omega')\,.
\ee
We stress that this expression, just like eq.~(\ref{eq:coll_prod}), does not rest on any 
assumptions about the {\it actual} DM distribution (other than being in the freeze-in regime).

The general expectation is that the contribution from decays in eq.~\eqref{C_inverse_decay} should 
be added to the contribution from $2\to2$ processes in eq.~\eqref{eq:boltzmann}. However, special 
care must be taken when both types of processes describe the same physical situation, such that 
adding the two would overestimate the DM production rate. This is the case in particular if the 
particle $A$ can also decay into two bath particles, $A \to \psi\psi$. The process 
$\chi\chi\to A^*\to \psi\psi$ then receives a resonant enhancement for $\sqrt{s} \approx m_A$, 
corresponding to the production of an on-shell mediator that subsequently decays into a pair of bath 
particles. Ref.~\cite{DeRomeri:2020wng} proposes to address this issue by cutting out the resonant 
region in the $2\to2$ process, such that the decay contribution can be consistently added. Here we 
will show that it is possible to instead consistently include the decay contribution in the $2\to2$ 
process by adopting an appropriate prescription for the Breit-Wigner propagator. A similar approach 
was advocated previously in ref.~\cite{Belanger:2018ccd}, but we provide additional physical insight 
on why such a prescription is plausible.

Let us consider the case where the dominant processes changing the comoving mediator number 
density  are decays such as $A \to \psi \psi$ and inverse decays such as $\psi \psi \to A$, i.e.\ we 
assume that the rates of other processes such as $\psi \psi \to A \psi'$ are negligible. We furthermore 
assume that the total decay width $\Gamma_\text{tot}$ of $A$ is larger than the Hubble rate, such 
that the mediator is brought into thermal equilibrium with the heat bath. Note that this assumption 
also implies that the partial width for decays into DM particles, $\Gamma_{A \to \chi\chi}$, which 
must be much smaller than the Hubble rate in the freeze-in regime, only gives a negligible 
contribution to the total width. 

For a scalar resonance $A$, the averaged amplitude squared for the annihilation process 
$\chi\chi\to A^*\to \psi\psi$ can be written as\footnote{%
We note that eq.~(\ref{eq:res_decomposition}) no longer holds as an {\it equality}
for vector resonances $A$ -- but can still be used in the form of a {\it replacement} 
when calculating the total cross section in vacuum, i.e.~eq.~(\ref{sigma_def}) without
quantum correction factors.
In general, however, this replacement is only valid if spin correlations can be neglected.
For a more detailed discussion see, e.g., ref.~\cite{Bringmann:2017sko}.
}
\be
\label{eq:res_decomposition}
 \overline{|{\cal M}_{\chi\chi\to \psi\psi}|^2} =
 \frac{ \overline{|{\cal M}_{\chi\chi\to A^*}|^2} ~\overline{|{\cal M}_{A^*\to \psi\psi}|^2}}{(s-m_A^2)^2+m_A^2\Gamma_{\rm BW}^2}\,,
\ee
where $\Gamma_{\rm BW}$ is the total width of $A$ as it appears in the Breit-Wigner propagator.
If kinematically accessible, the mediators will be dominantly produced on-shell and we
can adopt the narrow width approximation (NWA), 
\be
\label{eq:NWA}
\frac{1}{(s-m_A^2)^2+m_A^2\Gamma_{\rm BW}^2}
\to \frac{\pi}{m_A\Gamma_{\rm BW}}\delta(s-m_A^2)=\frac{\pi}{2m_A \omega'\Gamma_{\rm BW}}\delta(E+\tilde E-\omega')\,.
\ee
 Assuming furthermore $CP$ symmetry, 
implying $|{\cal M}_{\chi\chi\to A}|^2=|{\cal M}_{A\to\chi\chi}|^2$,
the collision term for annihilations, eq.~(\ref{eq:coll_prod}), thus becomes
\bea
\label{eq:C_ann_res}
C_{\rm ann}[f_\chi]&=&\frac{1}{N_\chi}\int\!\!\frac{d^3 p}{(2\pi)^3 2E}\int\!\!\frac{d^3\tilde p}{(2\pi)^32\tilde E}
  \left|\mathcal{M}\right|^2_{A\rightarrow\chi\chi} f_\chi^{\rm MB}(E) f_\chi^{\rm MB}(\tilde E)
  \frac{\pi}{\omega'}\delta(\omega'-E-\tilde E)\nonumber\\
  &&\times\frac{\Gamma_{\psi\psi}}{\Gamma_{\rm BW}}
  \overline{G}_{\psi\psi}(\gamma,m_A^2)\,,
\eea
where $\overline{G}$ is defined in eq.~(\ref{eq:gbar_def}) and 
\be
 \Gamma_{\psi\psi}=
 \frac{1}{2 m_A}
\int\!\!\frac{d^3k}{(2\pi)^32\omega}\int\!\!\frac{d^3\tilde k}{(2\pi)^32\tilde \omega}\,\delta^{(4)}(\tilde p+p-\tilde k-k)
\left|\overline{\mathcal{M}}\right|^2_{A\to\psi\psi}
\ee
is the standard partial decay width for $A\to\psi\psi$.

In the NWA, with all mediators created on-shell, we expect eq.~(\ref{eq:C_ann_res}) and
eq.~(\ref{C_inverse_decay}) to agree. This implies that the Breit-Wigner width for
a mediator in thermal equilibrium must in general be chosen as
\be
\label{eq:width_prescription}
\Gamma_{\rm BW}=\frac{1}{1+f_A(\omega')}\sum_{\psi_1\psi_2}\Gamma_{\psi_1\psi_2}\overline{G}_{\psi_1\psi_2}(\gamma,m_A^2)\,,
\ee
where the sum runs over all relevant heat bath particle $\psi_i$.
In fact, the origin of the additional terms (compared to the total decay width in vacuum) is 
straight-forward to understand: {\it (i)} the factor of $\overline{G}_{\psi_1\psi_2}$ modifies the
partial decay rate in vacuum, $\Gamma_{\psi_1\psi_2}$, such as to include the effect of Bose 
enhancement or Pauli blocking in the final state plasma particles; {\it (ii)} the overall
suppression factor of $1/(1+f_A)$ is a direct consequence of the fact that the imaginary part of the 
mediator self-energy at finite temperature is not given by the total decay rate, but rather 
by the difference between decay and inverse decay rates~\cite{Weldon:1983jn}.

The prescription for the mediator width in the $s$-channel given in eq.~\eqref{eq:width_prescription} 
ensures that the contribution from decay (of the same mediator from the thermal bath) is 
automatically accounted for in the collision term for $2\to2$ processes. In this case it would be 
inconsistent to add the collision term for the decay process, which would lead to a double-counting. 
In the following, we will therefore exclusively consider the $2\to2$ process (with the prescription 
discussed above) to calculate the DM production rate. 

We note that if there are processes that are not in equilibrium (i.e.\ they proceed dominantly in one 
direction, such as decays into ``invisible'' particles), the distribution of $A$ will be slightly different 
from an equilibrium distribution. This can be accounted for in eq.~(\ref{C_inverse_decay})  by 
rescaling with a factor $\Gamma_\text{eq} / (\Gamma_\text{eq} + \Gamma_\text{non-eq})$, where 
$\Gamma_\text{eq}$ ($\Gamma_\text{non-eq}$) denotes the total rate of all processes that are (are 
not) in thermal equilibrium. At the same time, $\Gamma_\text{non-eq}$ needs to be added to 
$\Gamma_\text{BW}$, leading once again to agreement between the two approaches. 
It is plausible that the same prescription as in eq.~\eqref{eq:width_prescription} can be used also 
when the mediator width is large such that the NWA becomes inaccurate, and when additional 
processes contribute to the thermalisation of the mediators. However, since we will not encounter 
such situations in the models that we study below, we will not explore these interesting directions, 
which may be more appropriately studied in a fully quantum field theoretical 
approach~\cite{Laine:2016hma,Jackson:2021dza}.

To conclude this discussion, we emphasise again that there are many situations where it \emph{is} 
necessary to include eq.~\eqref{C_inverse_decay} explicitly. This is the case for example when the 
mediator cannot decay into bath particles (e.g. because all such decays are kinematically forbidden) 
or if such decays are not sufficient to thermalise the mediator. For a recent discussion of such a 
set-up, we refer to ref.~\cite{Biondini:2020ric}.

\section{Finite-temperature effects}
\label{sec:finiteT}

In the previous section we have shown that in-medium effects can be straight-forwardly included in our formulation of the freeze-in formalism, such that interaction rates can be easily calculated for given particle masses and interactions. The remaining challenge is then to understand how the masses and interactions themselves depend on temperature. This is particularly relevant for quantities that depend on the Higgs vev, which varies strongly with temperature and vanishes for temperatures above the electroweak phase transition (section \ref{subsec:Veff}).
On top of that, thermal masses are generated from interactions with the plasma 
(section \ref{subsec:mT}). Also the QCD phase transition plays an important role as it changes 
fundamentally the relevant degrees of freedom that enter into our calculations 
(section \ref{subsec:qcd}).

In the remainder of this work we will use the expression ``finite-temperature effects'' to refer to temperature dependent masses and vevs as well as phase transitions. In contrast, the expression ``in-medium effects'' refers to both finite-temperature effects and quantum statistics.

\subsection{The effective Higgs potential}
\label{subsec:Veff}

The one-loop effective Higgs potential at finite temperature can be written as~\cite{Quiros:1999jp}
\begin{align}
V_\mathrm{eff}(\phi_c, T) = V_0(\phi_c) + V_1(\phi_c) + V_T(\phi_c, T) \, ,
\label{eq:veff_full}
\end{align}
where $\phi_c$ is the constant background field, and $V_0$, $V_1$ and $V_T$ represent the zero-temperature tree- 
and loop-level potential, and the finite temperature potential, respectively:
\begin{align}
V_0(\phi_c) &= -\frac{m^2}{2} \phi_c^2 + \frac{\lambda}{4}\phi_c^4\,,\\
V_1(\phi_c) &= -\frac{1}{64\pi^2}\sum_{i=W, Z, t}\varepsilon_i n_i\left(m_i^4(\phi_c)\left(\log\frac{m_i^2(\phi_c)}{m_i^2(v)} - \frac{3}{2} \right) + 2 m_i^2(v)\,m_i^2(\phi_c)\right)\,,\\
V_T(\phi_c, T) &= -\frac{T^4}{2\pi^2}\sum_{i=W,Z,t}(\varepsilon_i n_i) \int_0^\infty x^2 \log(1-\varepsilon_i e^{-\sqrt{x^2 + m(\phi_c)^2/T^2}})\, dx\, \label{eq:veff_temp}.
\end{align}
In these equations, $m_i (\phi_c)$ are field-dependent masses and $m_i(v)$ their zero-temperature values. The factors 
$n_i$ correspond to the total degrees of freedom for a particle species ($n_W = 6$, $n_Z=3$, and $n_t = 12$), and 
$\varepsilon_i = 1\,(-1)$ for fermions (bosons) as before. At sufficiently large temperatures, we can expand the integrals 
in eq.~(\ref{eq:veff_temp}) in powers of $m^2(\phi_c)/T^2$. Ignoring the field-independent terms, we can then write the 
total effective potential to leading order as
\begin{align}
\label{eq:Vhightemp}
V_\mathrm{eff, high-T} = D (T^2 - T_0^2) \phi_c^2 - E T \phi_c^3 + \frac{\lambda(T)}{4}\phi_c^4\,,
\end{align}
with the parameters $D$, $E$ and $T_0$,  as well as $\lambda(T)$, provided in ref.~\cite{Quiros:1999jp}. 

Using this expansion, the critical temperature of the electroweak phase transition is evaluated to be 
at $T_{\rm EW} \sim 163 \,\mathrm{GeV}$.  For $T > T_{\rm EW}\approx T_0$ the 
effective potential thus has a global minimum at $\phi_c = 0$, with the thermal mass of the complex Higgs doublet 
given by
\begin{align}
\label{eq:mhunbroken}
m_H^2(T)=\frac{d^2 V_\mathrm{eff}}{d\phi_c^2}\Bigr|_{\phi_c=0} \simeq 2D(T^2 - T_0^2)\,.
\end{align} 
For $T\lesssim T_0$ a second minimum appears in the effective potential, relaxing shortly thereafter, for 
$T\leq T_{\rm EW}$, to a global minimum. 
Using the high-temperature expression above, it is possible to also derive analytic estimates for the temperature 
dependence of the (physical) Higgs mass and vev after the phase transition~\cite{Quiros:1999jp}, although these 
expressions necessarily break down for $T\ll T_{\rm EW}$. 
To ensure that the correct zero-temperature Higgs mass is reproduced, therefore, we numerically minimise 
the full expression for $V_\mathrm{eff}$  given in eq.~(\ref{eq:veff_full}) to obtain $m_h(T)$ and $v(T)$. 

\begin{figure}
\centering
\includegraphics[width=0.49\textwidth, clip, trim = 15 15 15 15]{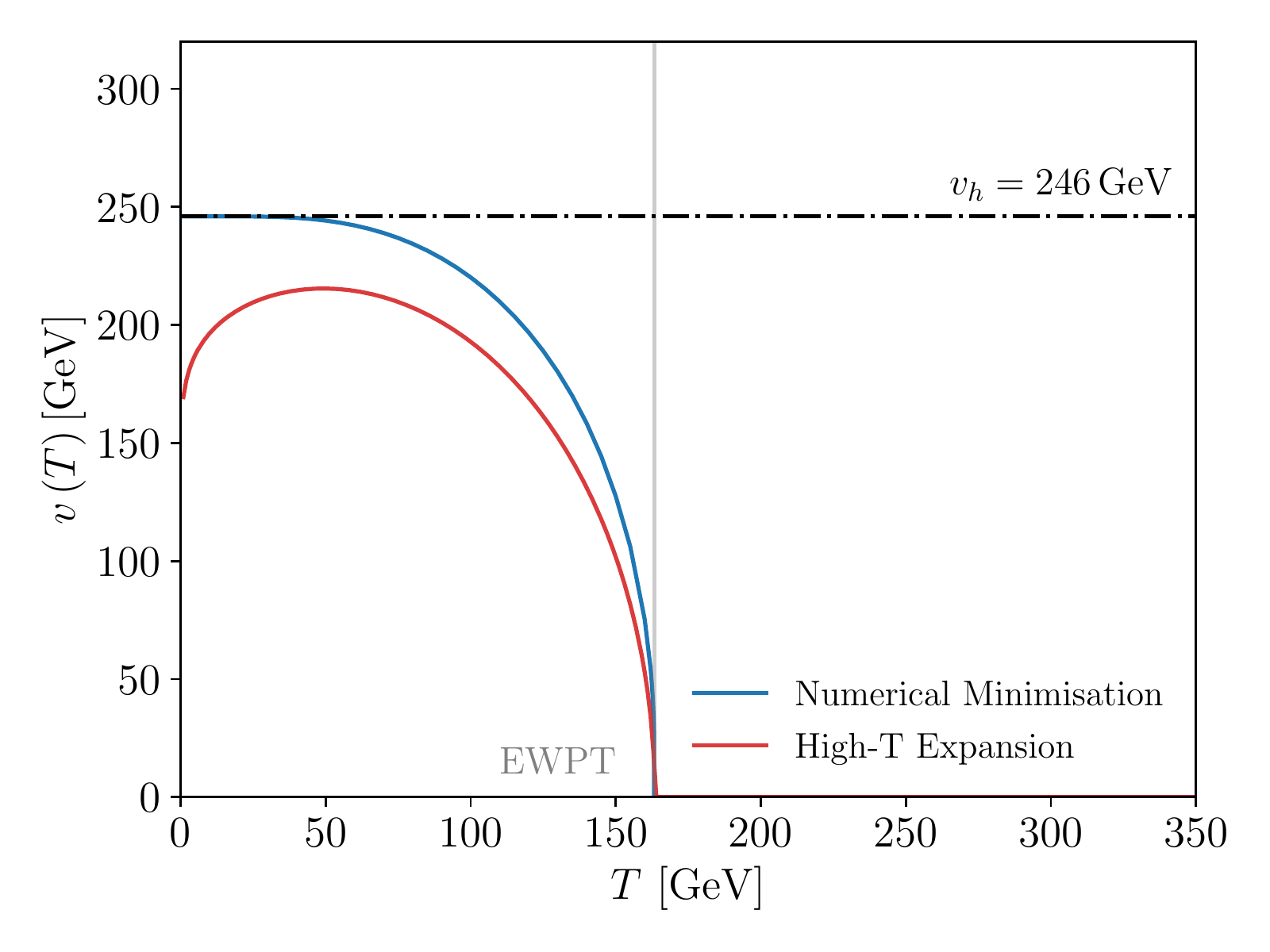}
\includegraphics[width=0.49\textwidth, clip, trim = 15 15 15 15]{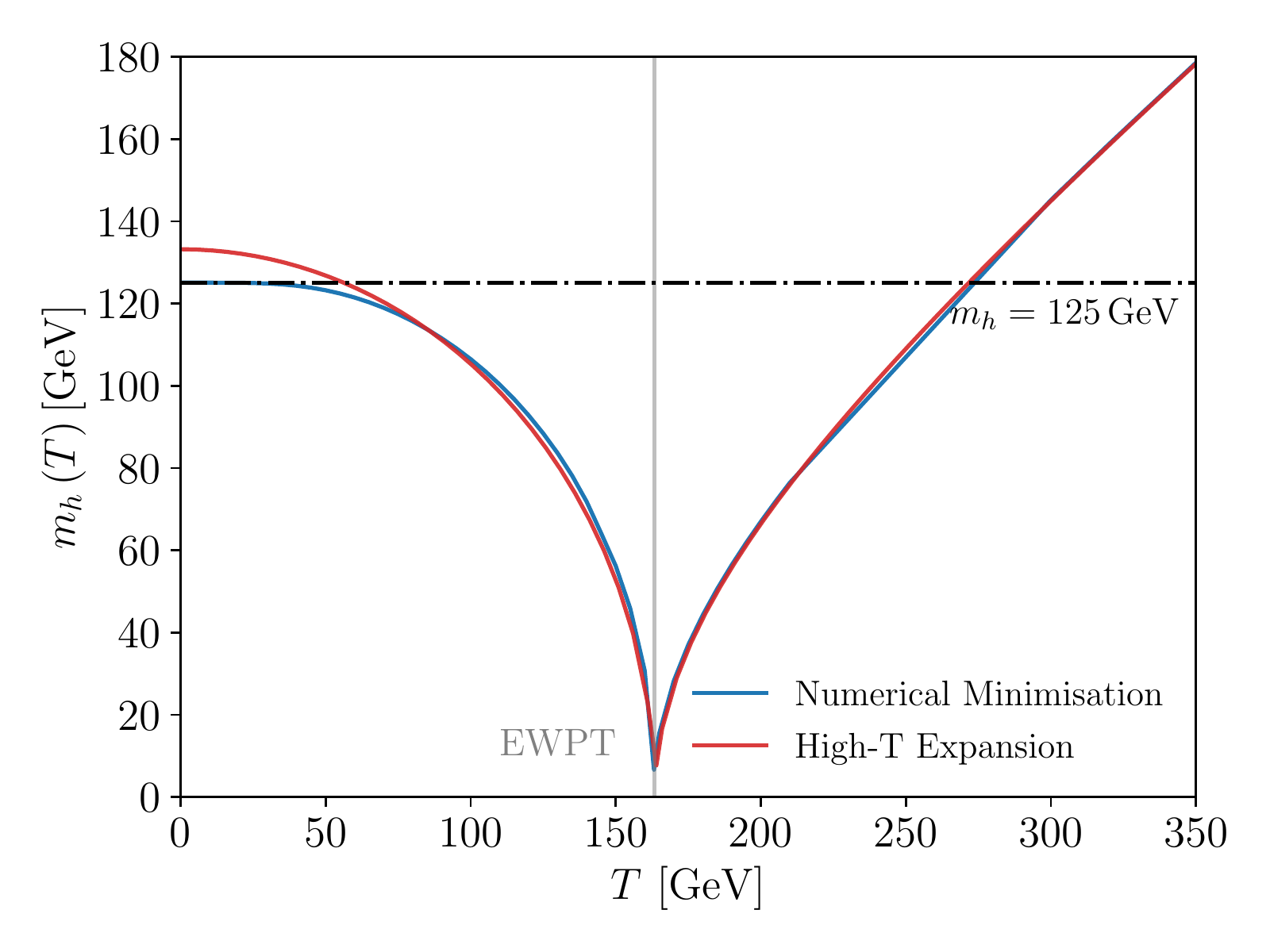}
\caption{\label{fig:vev_temp}%
The Higgs vev (left) and mass (right) as a function of temperature. The red lines 
correspond to results obtained by using a high-temperature expansion of the effective potential, whereas the blue lines 
correspond to the ones used in this work, obtained by numerical minimisation of eq.~(\ref{eq:veff_full}). 
The dashed-dotted black lines represent the zero temperature values of the Higgs vev (left) and mass (right). The 
vertical grey line indicates the approximate temperature of the electroweak phase transition.}
\end{figure}

A comparison between our numerical results and the analytic estimates for the Higgs vev and mass is shown in 
figure~\ref{fig:vev_temp}. 
The breakdown of the analytic estimate at low temperatures is clearly visible, in particular in the left panel
of the figure, where a logarithmic divergence in the analytic expression for $\lambda(T)$ leads to $v\to0$ in this limit. 
The numerical results that we use in this work, on the other hand, correctly reproduce the 
zero temperature values for $m_h$ and $v_h\equiv v(T=0)$.
Both prescriptions lead to very similar estimates of the critical temperature, $T_{\rm EW}$, 
and the agreement is generally very good at high temperatures. 
Let us however point out that in our approximation for the effective potential, we have neglected the contributions 
from so-called ring or daisy diagrams~\cite{Curtin:2016urg,Quiros:1999jp} 
which regulate the IR divergences in the theory especially close to the 
temperature of the phase transition. This is the main origin of the small but finite value of  
$m_h(T_{\rm EW})\approx 10\,\text{GeV}$ that is visible in the right panel of figure~\ref{fig:vev_temp}. 
We note that a full determination of the effective potential is anyway beyond the scope of this
work, likely involving lattice calculations~\cite{DOnofrio:2015gop}, and that the exact behaviour of $m_h$ and $v$
very close to the phase transition has a negligible impact on our results.
In practice, to avoid numerical instabilities at the electroweak phase transition (EWPT), we smooth out the reaction rates over a 
small window around $T_\text{EW}$.

\subsection{Thermal masses}
\label{subsec:mT}
Interactions in the plasma also modify the dispersion relations of SM particles, thereby generating effective mass terms 
both before and after the electroweak phase transition. For $T<T_{\rm EW}$, the masses acquired via the Higgs 
mechanism dominate, except for temperatures very close to the phase transition where the two contributions
may be comparable and hence need to be added (in quadrature for bosons). For $T>T_{\rm EW}$, however, gauge 
bosons and fermions only have thermal masses generated as a result of screening 
effects in the plasma. For gauge bosons, the effective thermal mass at leading order is given by~\cite{Rychkov:2007uq}
\begin{align}
m^2_V = \frac{1}{6} g^2 T^2 (N + N_S + \frac{N_F}{2})\,,
\end{align}
where $g$ denotes the gauge coupling and the coefficients $N,\,N_F$ and $N_S$  
parameterise the 1-loop contributions from 
vectors, fermions and scalars charged under the given gauge group. In the SM, 
$N=3,\,N_F = 6,\, N_S=0$ for $SU(3)_c$, $N=2,\,N_F = 6,\, N_S=1/2$ for $SU(2)_L$ and $N=0,\,N_F = 10,\, N_S=1/2$ 
for $U(1)_Y$. 
More precisely, the above expression corresponds to the transverse mass, which for a relativistic 
gauge boson is larger than the plasma frequency by a factor $3/2$~\cite{Raffelt:1999tx}. 

Fermions before the electroweak phase transition are chiral, with left- and right-handed particles 
having different effective mass terms. For leptons, the thermal masses receive contributions from $SU(2)_L$, $U(1)_Y$ and Yukawa 
couplings, and can be written as~\cite{Elmfors:1993re}
\bea
m_{l, \mathrm{L}}^{2}&=&\frac{m_{Z}^{2}+2 m_{W}^{2}+m_{l}^{2}+m_{l^{\prime}}^{2}}{2 v_{h}^{2}} T^{2}\,, \\
m_{l, \mathrm{R}}^{2}&=&\frac{m_{Z}^{2}-m_{W}^{2}+\frac{1}{2} m_{l}^{2}}{2 v_{h}^{2}} T^{2}\,,
\eea
where we made use of tree-level relations between $SU(2)_L\times U(1)_Y$ couplings
and gauge boson masses, after spontaneous symmetry breaking, 
and $m_{l,\,l^\prime}$ are the zero-temperature masses of the leptons in an $SU(2)$ doublet. 
For quarks, also strong interactions need to be taken into account, such that the corresponding expressions 
become 
\bea
m_{q, \mathrm{~L}}^{2}&=&\frac{1}{6} g_{s}^{2} T^{2}+\frac{3 m_{W}^{2}+\frac{1}{9}\left(m_{Z}^{2}-m_{W}^{2}\right)+m_{u}^{2}+m_{d}^{2}}{8 v_{h}^{2}} T^{2}\,, \\
m_{u, \mathrm{R}}^{2}&=&\frac{1}{6} g_{s}^{2} T^{2}+\frac{\frac{4}{9}\left(m_{Z}^{2}-m_{W}^{2}\right)+\frac{1}{2} m_{u}^{2}}{2 v_{h}^{2}} T^{2}\,, \\
m_{d, \mathrm{R}}^{2}&=&\frac{1}{6} g_{s}^{2} T^{2}+\frac{\frac{1}{9}\left(m_{Z}^{2}-m_{W}^{2}\right)+\frac{1}{2} m_{d}^{2}}{2 v_{h}^{2}} T^{2}\,.
\eea
After the phase transition, the thermal corrections to the electroweak fermion masses are 
approximately given by
\bea
\Delta m_{l}^{2}&=&\frac18{e^{2} T^{2}}, \\
\Delta m_{q}^{2}&=&\frac16{g_{s}^{2} T^{2}}\,,
\eea
where we neglected electromagnetic contributions to the quark masses.

We note that, for the specific application explored in Section \ref{sec:scalarsinglet}, we are mostly 
sensitive to the thermal Higgs mass at temperatures above the electroweak phase transition,
as displayed in the right panel of figure~\ref{fig:vev_temp}. We still implement the above expressions
for all SM particles, and make them available as general convenience functions in \ds\ (see also 
appendix \ref{app:ds}).

\subsection{The QCD phase transition}
\label{subsec:qcd}
The QCD phase transition is linked to the breaking of the chiral symmetry and the formation of a 
quark condensate in vacuum, $\langle q \bar{q}\rangle$. This symmetry breaking results in the 
confinement of quarks in colour-neutral bound states at low energies. The relevant degrees of 
freedom are therefore baryons and mesons (rather than free quarks and gluons), which can be 
described using an effective field theory approach, i.e.\ chiral perturbation theory. A detailed 
modelling of the QCD phase transition is highly challenging, but to first approximation we expect the 
phase transition to occur when the temperature drops below the confinement scale 
$\Lambda_\text{QCD}$. In the following we will use a fiducial value of 
$T_\text{QCD} = 154 \, \mathrm{MeV}$, referring to Refs.~\cite{Schwarz:2003du,Husdal:2016haj} for 
a more detailed discussion (and noting that more recent QCD lattice simulations tend to prefer 
slightly higher values of up to $\sim158$\,MeV~\cite{HotQCD:2018pds,Borsanyi:2020fev}). 

In our context, the QCD phase transition manifests itself in two ways.
First, it directly affects the effective number of relativistic degrees of freedom in the heat bath of 
the early universe, entering in the quantity $\tilde H$ in eq.~(\ref{eq:dYdx}). For this we update the 
default prescription of \ds, to incorporate results from lattice simulations as well as perturbative 
computations up to the 3-loop level~\cite{Laine:2006cp,Laine:2015kra}.
Second, for low temperatures and CMS energies $\sqrt{s}\lesssim2$\,GeV, the DM annihilation cross section can 
generally no longer be approximated by assuming free quarks in the final state. Instead, the
cross section must be calculated within the framework of chiral perturbation theory to adequately 
take into account the hadronic nature of the final states. We will discuss this in more detail in 
the specific context of off-shell Higgs decays, which is the topic of the following section.

\section{Off-shell Higgs decays}
\label{sec:hdecay}

For a broad class of DM models, annihilation into SM final states proceeds via the so-called Higgs portal, i.e.\ via an 
off-shell Higgs boson that can decay into the various SM fermions and gauge bosons.\footnote{A second well-studied class of DM models introduces a new scalar mediator that couples to SM particles through mixing with the SM Higgs boson. Although we will not discuss these models here, our results can be directly applied to this case as well.} The total annihilation cross 
section is then directly proportional to the off-shell Higgs width 
$\Gamma_{h^\ast}(\sqrt{s}) \equiv \Gamma_h(m_h = \sqrt{s})$, where $\sqrt{s}$ denotes the CMS energy. 
It therefore becomes essential to have an accurate calculation of this off-shell 
width, for arbitrary values of $\sqrt{s}$ and for finite temperatures. We stress that, as discussed in 
Section \ref{sec:freeze-in},  such an improved estimate of the DM annihilation rate can also be used to 
calculate the DM {\it production} in freeze-in scenarios, to a corresponding degree of accuracy.

We begin with a discussion of the relevant Higgs decay modes 
and their implementation for $\sqrt{s}\gtrsim 2$\,GeV 
at zero temperature, and how to avoid  unitarity violation in the limit where $\sqrt{s} \gg v_h$.
We then consider the impact of finite temperature effects on these considerations.
Finally, we discuss the case of $\sqrt{s} < 2 \, \mathrm{GeV}$ and the peculiarities of the QCD phase transition.

\subsection{Relevant decay modes at zero temperature}
\label{sec:tree_decays}

For $2 \, \mathrm{GeV} \lesssim \sqrt{s} \lesssim 1 \, \mathrm{TeV}$ the off-shell Higgs decay width can be calculated perturbatively~\cite{Cline:2013gha}. The tree-level decay widths into fermions are given by
\begin{equation}
  \Gamma(h\to f \bar{f}) = \frac{N_c m_f^2 \sqrt{s}}{8\pi v_h^2} \left(1 - 4 \eta_f\right)^{3/2} , 
\end{equation}
where $\eta_X \equiv m_X^2 / s$ and $N_c = 1$ (3) for leptons (quarks). For decays into two real gauge bosons $V = W,Z$ one finds
\begin{equation}
  \Gamma(h\to V V) = \frac{\sqrt{s}^3}{32 v_h^2 \pi} \delta_V \sqrt{1-4\eta_V}(1-4\eta_V+12 \eta_V^2)
  \label{eq:htoVV}
\end{equation}
with $\delta_W = 2$ and $\delta_Z = 1$. The leading-order decay width into gluons is given by~\cite{Djouadi:2005gi}
\begin{equation}
 \Gamma(h\to gg) = \frac{\alpha_s^2(\sqrt{s}) \sqrt{s}^3}{72 v_h^2\, \pi^3}\Bigr| \frac{3}{4}\sum_q A_{1/2}^h\left(\tfrac{1}{4\eta_q}\right)\Bigr|^2\,,
\end{equation}
where $\alpha_s(\sqrt{s})$ denotes the running strong coupling and 
\begin{equation}
A_{1/2}^h(\tau) = 2 (\tau + (\tau-1)f(\tau))\tau^{-2}
\end{equation}
with
\begin{align}
f(\tau) = 
\begin{cases}
\arcsin^2\sqrt{\tau} & \tau \leq 1 \\
-\frac{1}{4}\left(\log\frac{1+\sqrt{1-\tau^{-1}}}{1-\sqrt{1-\tau^{-1}}}-i\pi\right)^2 & \tau>1 \, .
\end{cases}
\end{align}
The decay width into photons can be written in an analogous way but gives a negligible contribution to the total decay width.

In practice, higher-order corrections are non-negligible. This is particularly true for $\sqrt{s} \gg v_h$, 
where the decay 
into gauge bosons dominates and large NLO EW corrections arise from the (almost) on-shell 
emission of additional gauge bosons, as well as close to final state thresholds. 
To capture these and other effects, we use the tabulated decay widths from \hd~\cite{Djouadi:2018xqq} up to 
$\sqrt{s} \sim 1 \, \mathrm{TeV}$. For even larger CMS energies, additional 
modifications become necessary, which will be discussed next.

\subsection{Unitarization}

\begin{figure}[t]
	\centering
	\includegraphics[width=0.6\textwidth, clip, trim = 1 1 10 10]{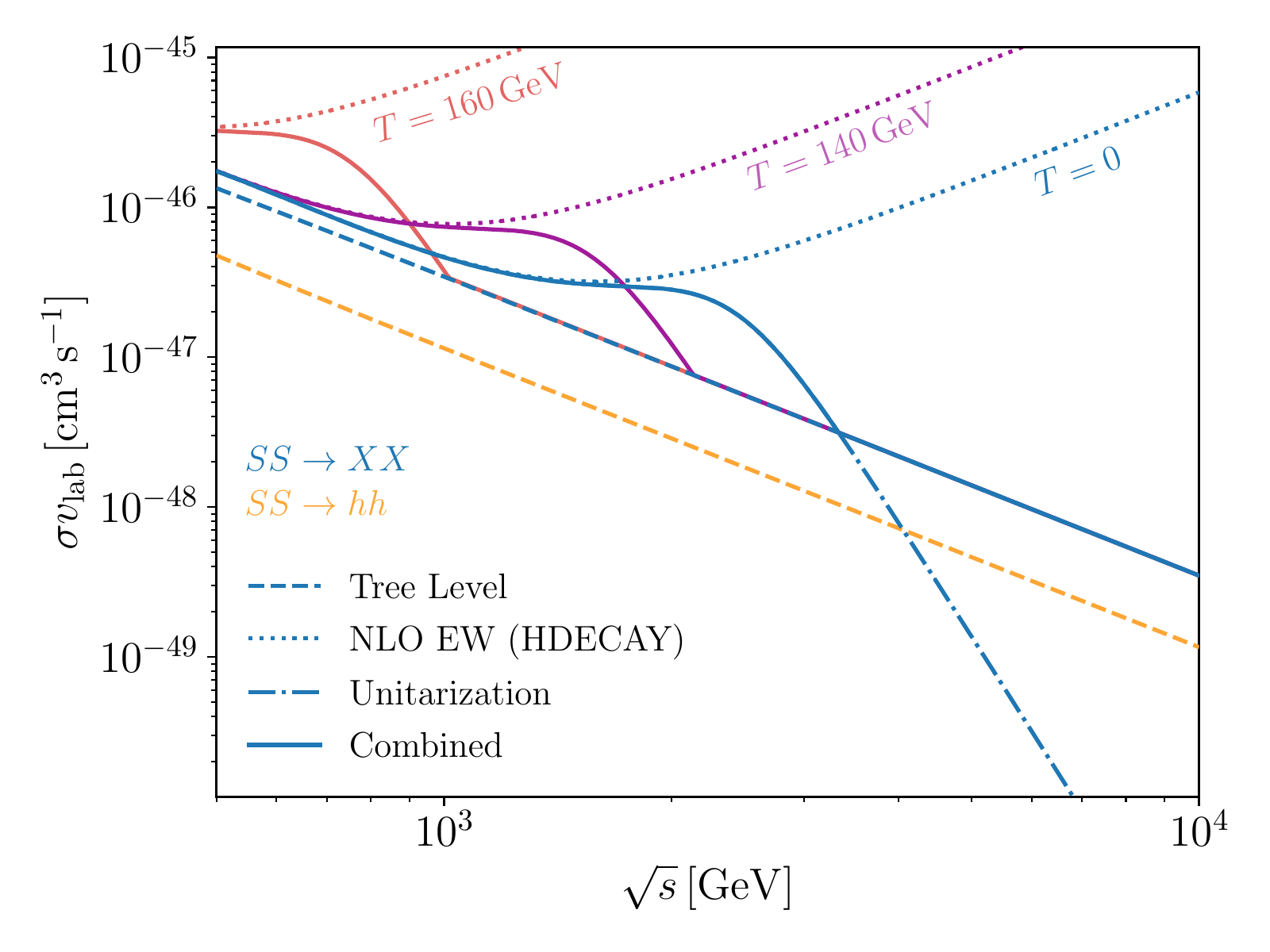}
	\caption{Total annihilation cross section for the toy model introduced in eq.~(\ref{eq:toy}), for $\lambda_{hs} = 10^{-11}$.
	The solid line shows the prescription that we adopt in this work, as stated in eq.~(\ref{eq:SS_XX_mod}), 
	thus modifying the result based on the Higgs decay width returned by \hd\ (dotted line) such
	as to avoid unitary violation and falling back to the tree-level result (dashed line) at high CMS energies. Different colours show these results in the zero-temperature limit (blue), as well
	as for $T=140\,\mathrm{GeV}$ (purple) and $T=160\,\mathrm{GeV}$ (red).
	For 	comparison, the orange dashed line shows the full tree-level cross section into a pair of SM Higgs bosons.}
	\label{fig:unitarization}
\end{figure}

For $\sqrt{s} \gg 1 \, \mathrm{TeV}$ the off-shell decay width returned by \hd\ becomes unphysical. To see this, it is 
helpful to consider a model in which the SM Higgs boson is coupled to a real scalar singlet $S$ via
\begin{equation}
\label{eq:toy}
 \mathcal{L} \supset \frac{\lambda_{hs}}{2} |H|^2 S^2 \, .
\end{equation}
This interaction is identical to the one of the scalar singlet DM model that will be discussed in more detail in 
Section \ref{sec:scalarsinglet}. The full annihilation cross section into SM Higgs bosons is, to leading order in 
$\lambda_{hs}$, provided in Refs.~\cite{Cline:2013gha,Bringmann:2018lay};
for $\sqrt{s} \gg v_h$, it simplifies to
\begin{equation}
\label{eq:SS_hh}
 \sigma(SS \to hh) v_{\rm lab} = \frac{\lambda_{hs}^2}{32 \pi s} \, ,
\end{equation}
where $v_{\rm lab}= \sqrt{s(s-4m_S^2)}/(s-2 m_S^2)$ with $m_S$ denoting the singlet mass at zero temperature.
The summed annihilation cross section into all other SM particles $X \neq h$, on the other hand, 
can be written as~\cite{Cline:2013gha}
\begin{equation}
\label{eq:SS_XX}
 \sigma(SS \to XX) v_{\rm lab} = \frac{\lambda_{hs}^2 v_h^2}{\sqrt{s}} \frac{1}{(s - m_h^2)^2 + m_h^2 \Gamma_{h}^2} \Gamma_{h^\ast}(\sqrt{s}) \, .
\end{equation}
These cross-sections are plotted in figure \ref{fig:unitarization} as a function of the CMS energy.
It is worth noting that eq.~(\ref{eq:SS_XX}) is valid to leading order in $\lambda_{hs}$, describing 
annihilation via an $s$-channel Higgs exchange, but fully encapsulates 
higher-order corrections in SM couplings.
Unitarity requires that for $s \to \infty$ any cross section falls at least as fast as $1/s$. 
For the specific process we are interested in we even expect $\sigma \propto 1/s$ at large centre-of-
mass energies from dimensional analysis, which implies that $\Gamma_{h^\ast}(\sqrt{s})$ should 
grow as $s^{3/2}$. This is the case for the tree-level decay width given in eq.~(\ref{eq:htoVV}). 
Indeed, using the tree-level expressions, one finds that in the limit $s \to \infty$ the cross section 
agrees with the cross section obtained in the limit $v_h \to 0$, i.e.\ when electroweak symmetry is 
restored:
\begin{equation}
\label{eq:equivalence}
 \sigma(SS \to hh) + \sigma(SS \to XX)^\text{tree} \to \sigma(SS \to HH)\,,
\end{equation}
with $ \sigma(SS \to HH)=4\times \sigma(SS \to hh)$ in this limit, 
as required by the Goldstone boson equivalence theorem.

However, the NLO EW corrections implemented in \hd\ predict 
a more rapid growth of $\Gamma_{h^\ast}(\sqrt{s})$, represented by the dotted lines in 
figure~\ref{fig:unitarization}. This means that even higher-order effects become increasingly 
important for large $\sqrt{s}$ in order to restore unitarity. Rather than attempting to compute these corrections explicitly, 
we will choose a phenomenological prescription to ensure that unitarity is never violated in our numerical 
implementation. For this we follow the approach from ref.~\cite{Khoze:2017tjt} and replace the on-shell Higgs decay 
width in the propagator by the off-shell Higgs decay width, such that
\begin{equation}
\label{eq:SS_XX_mod_temp}
 \sigma(SS \to XX)^\text{unitary} v_{\rm lab} = \frac{\lambda_{hs}^2 v_h^2}{\sqrt{s}} \frac{1}{(s - m_h^2)^2 + m_h^2 \Gamma_{h^\ast}(\sqrt{s})^2} \Gamma_{h^\ast}(\sqrt{s}) \, .
\end{equation}
This approach is similar to the form factor unitarization approach from ref.~\cite{Arnold:2011wj}.

In the limit $s \to \infty$, however, this unitarised cross section decreases faster than $1/s$. 
This violates the above general argument from unitarity and dimensional analysis,
and is also inconsistent with the naive expectation that higher-order corrections should increase 
rather than decrease the decay rate (due to larger final-state multiplicities).
We therefore combine the two calculations as follows:
\begin{equation}
\label{eq:SS_XX_mod}
 \sigma(SS \to XX)^\text{total} = \text{max}\left[\sigma(SS \to XX)^\text{tree}, \sigma(SS \to XX)^\text{unitary}\right].
\end{equation}
The cross section defined in this way has the following desirable properties:
\begin{enumerate}
 \item It fully captures the effects of NLO EW corrections for $\sqrt{s} \lesssim \mathrm{TeV}$.
 \item There is no violation of unitarity for $\sqrt{s} \gtrsim \mathrm{TeV}$.
 \item Higher-order corrections do not decrease the tree-level result (of the broken theory)  for large CMS energies.
 \item The tree-level result in the unbroken theory is recovered in the limit $\sqrt{s} \to \infty$.
\end{enumerate}
The second point can be made more precise by considering the well-known unitarity bound on the DM annihilation cross 
section from ref.~\cite{Griest:1989wd}. While commonly quoted in the non-relativistic limit, it is straight-forward to 
generalise the calculation to relativistic DM particles, in which case the unitarity bound reads 
\begin{equation}
 \sigma_\text{ann} v_{\rm lab} < \frac{4 \pi}{\sqrt{s} \sqrt{s - 4 m_S^2}} \,.
 \label{eq:general_unitarity_bound}
\end{equation}
Applying this bound to $\sigma_\text{ann} = \sigma(SS \to XX)^\text{total}$ we obtain a bound on $\lambda_{hs}$ as a 
function of $\sqrt{s}$, which is most stringent for $\sqrt{s} \approx 2.2 \, \mathrm{TeV}$ and yields 
$\lambda_{hs} < 10.9$. This value should be compared to the tree-level bound $\lambda_{hs} < 8\pi$ first derived in 
ref.~\cite{Cynolter:2004cq} by considering the scattering process $S + h \to S + h$ in the limit 
$\sqrt{s} \to \infty$.%
\footnote{
We note that somewhat stronger bounds were recently obtained by considering finite values of $\sqrt{s}
$~\cite{Goodsell:2018tti}.
} 
Of course, the values of $\lambda_{hs}$ of interest in the context of freeze-in will be many orders of magnitude below 
this value.

\subsection{Finite-temperature corrections}

At first sight, the discussion of unitarity limits above may seem of limited practical relevance, given that the decay 
$h \to VV$ is only allowed after electroweak symmetry breaking (EWSB), at which point the temperature of the universe is so low that the probability 
for collisions with $\sqrt{s} \gg v(T)\approx v_h$ is exponentially suppressed. However, an analogous argument applies 
also for temperatures only slightly below the EWPT, where $v(T)\ll v_h$ allows for $\sqrt{s} \sim T \gg v(T)$.
To understand the temperature dependence of the off-shell Higgs decay width, we can express the masses of all SM 
fermions and gauge bosons through the Higgs vev, which is the only dimensionful quantity in the Standard Model at 
energies well above the QCD scale: $m_{f,V} \propto v$. Based on dimensional analysis,
and $\Gamma_{f,V} \propto m_{f,V} $, it then follows immediately that 
the analytical expressions for the partial decay widths given above can all be written as $\Gamma = v f(\sqrt{s}/v)$ with 
appropriate functions $f(x)$, i.e.\ they must depend on the CMS energy via the dimensionless ratio $\sqrt{s}/v$. 
We therefore conclude that the off-shell decay width at finite temperature is simply given by
\begin{equation}
\label{eq:Gamma_hT}
 \Gamma_{h^\ast}(T, \sqrt{s}) =  \frac{v(T)}{v_h} ~\Gamma_{h^\ast}\!\!\left(\sqrt{s} \frac{v_h}{v(T)}\right)
\end{equation}
in terms of the zero-temperature decay width $\Gamma_{h^\ast}(\sqrt{s})$.

As $T$ approaches the temperature of EWSB from below, $v(T) \to 0$ and hence $\sqrt{s} v_h/ v(T)$ diverges. The 
modification of the DM annihilation cross section at large $\sqrt{s}$ that we introduced above to avoid unitarity violation 
therefore also becomes relevant close to the EWPT (see also the red and purple lines in 
figure~\ref{fig:unitarization}). By construction this modification ensures that the limit $v(T) \to 0$ 
is smooth and converges to the annihilation cross section in the unbroken phase, 
cf.~eq.~(\ref{eq:equivalence}) above.

\subsection{Chiral symmetry breaking}

\hd\ in principle also allows for the calculation of the off-shell Higgs decay width for $\sqrt{s}$ as small as 
$1 \, \mathrm{GeV}$. However, it is implicitly assumed that the Higgs boson still decays into free quarks and gluons. 
This is a valid assumption for temperatures above the QCD phase transition, but it becomes inappropriate at smaller 
temperatures, where the confinement into hadrons must be taken into account. For 
temperatures below the QCD phase transition and $\sqrt{s} \lesssim2 \, \mathrm{GeV}$, the off-shell decay width into QCD 
bound states can instead be calculated in chiral perturbation theory with form factors obtained from dispersion 
relations~\cite{Winkler:2018qyg}. 

\begin{figure}
\centering 
\includegraphics[width=0.49\textwidth]{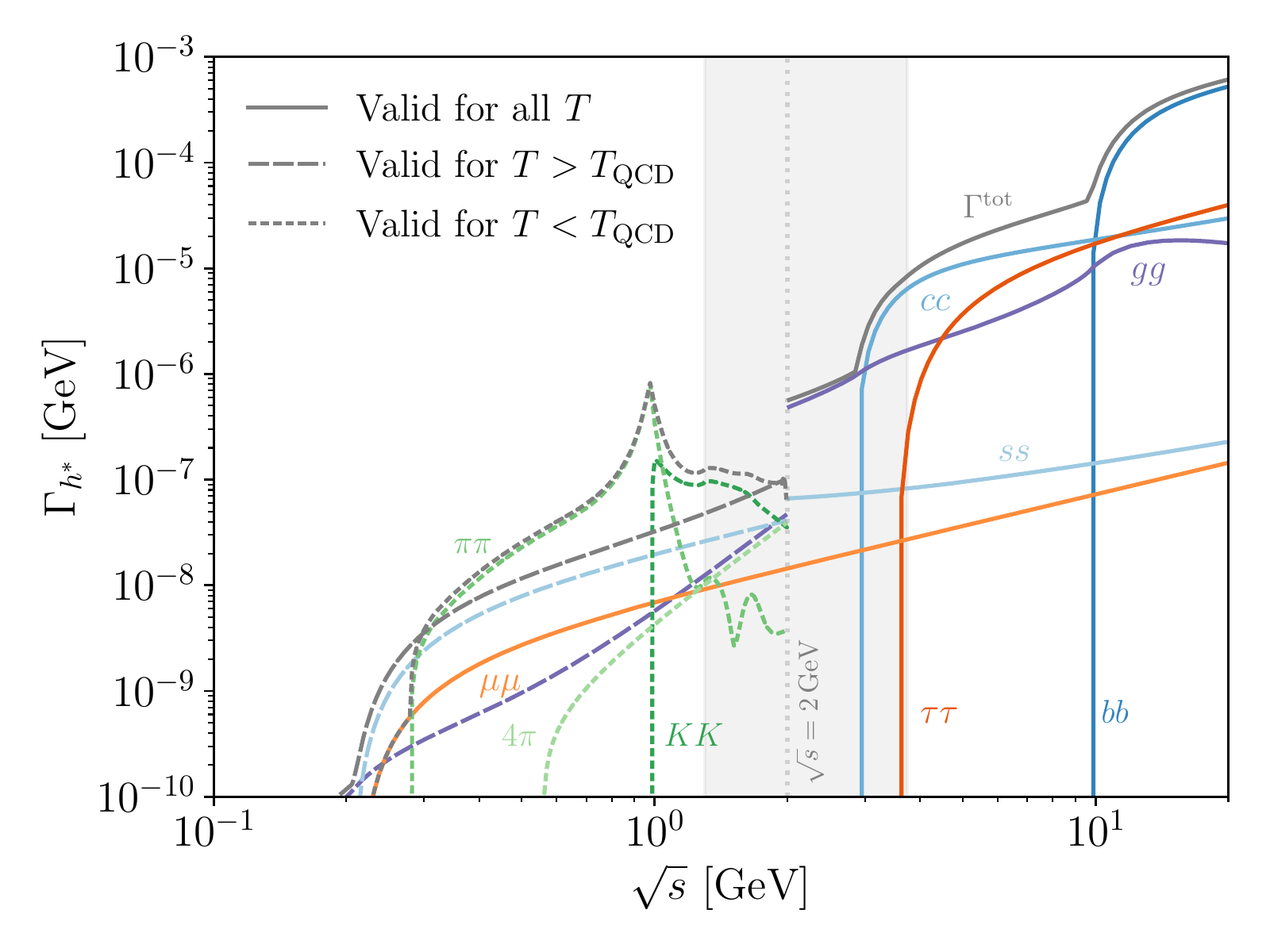}
\includegraphics[width=0.49\textwidth]{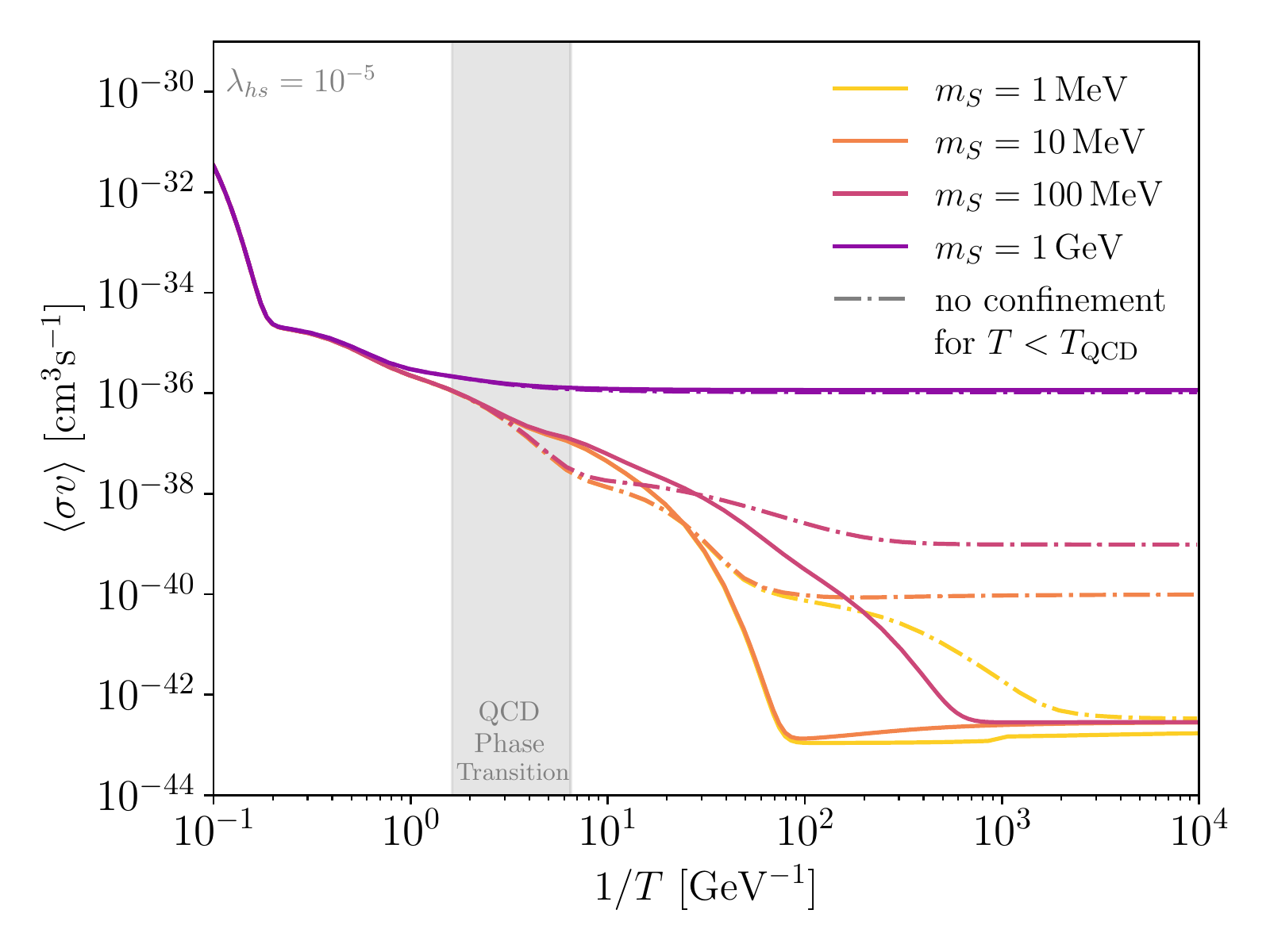}
\caption{{\it Left:} Partial decay widths of an off-shell SM Higgs boson as a function of the CMS 
energy. For $\sqrt{s} < 2 \, \mathrm{GeV}$ we use the decay widths into hadrons from ref.~\cite{Winkler:2018qyg} 
for $T < T_\mathrm{QCD}$ (short-dashed lines) and leading-order estimates 
of the decay widths into free quarks and gluons for $T > T_\mathrm{QCD}$ (long-dashed lines). For 
higher CMS energies we use the results from \hd~\cite{Djouadi:2018xqq} based on the assumption of free quarks 
and gluons in the final states and including higher-order corrections (solid lines). In the grey area the decay 
widths suffer from significant uncertainties due to the transition between the two regimes, see text for details.
{\it Right:} Thermally averaged annihilation cross section as a function of inverse temperature, including
quantum statistics and for DM masses as indicated. Solid lines show the result obtained when using hadronic 
final states for the off-shell Higgs decay widths for $T < T_\mathrm{QCD}$,
while dot-dashed lines result from (incorrectly) assuming decays into free quarks even below the QCD phase transition 
(indicated by the grey band).
}
\label{fig:gammah_updated}
\end{figure}

In the present work we therefore take the predictions from \hd\ for $\sqrt{s} > 2\,\mathrm{GeV}$ and those from 
ref.~\cite{Winkler:2018qyg} for $\sqrt{s} < 2 \, \mathrm{GeV}$ and $T < T_\mathrm{QCD}$. For $T > T_\mathrm{QCD}$ one could in 
principle continue using the results from \hd\ also for $\sqrt{s} < 2 \, \mathrm{GeV}$. However, we find that the 
higher-order corrections implemented in \hd\ become unreliable at such small values of $\sqrt{s}$ due to the strong 
coupling $\alpha_s$ becoming non-perturbative. For $T > T_\mathrm{QCD}$ and $\sqrt{s} < 2 \, \mathrm{GeV}$ we 
therefore simply use the leading-order results for decays into free quarks and gluons given in Section 
\ref{sec:tree_decays}. The adopted decay 
widths are illustrated in the left panel of figure~\ref{fig:gammah_updated}, with long-dashed lines corresponding to high
temperatures ($T > T_\mathrm{QCD}$) and short-dashed lines corresponding to low temperatures ($T < T_\mathrm{QCD}$). 
Note that for $\sqrt{s} > 2 \, \mathrm{GeV}$, as well as for leptons, we use the same prescription in both cases (solid lines).
The curves shown in the left panel of figure~\ref{fig:gammah_updated} exhibit significant discontinuities as 
$\sqrt{s} = 2 \, \mathrm{GeV}$, which are indicative of the substantial uncertainties in the various approximations made. 
For the decay widths into hadrons, ref.~\cite{Winkler:2018qyg} states that they should only be considered as rough estimates for 
$\sqrt{s} > 1.3 \, \mathrm{GeV}$. Conversely, the decay width into charm quarks obtained from \hd\ does not take into 
account the fact that at low temperatures a pair of charm quarks can only be produced for 
$\sqrt{s} > 2 m_D \approx 3.74 \, \mathrm{GeV}$. The grey shading in the left panel of figure~\ref{fig:gammah_updated} is intended to 
caution the reader about these uncertainties.\footnote{%
In the context of numerical relic density calculations with \ds\ we implement a linear interpolation
of the SM Higgs decay rate within the grey band,  
thus avoiding the (unphysical) discontinuity visible in the figure when computing interaction rates.
}

To complement this discussion we show in the right panel of figure~\ref{fig:gammah_updated} the total thermally 
averaged annihilation cross-sections $\langle \sigma v\rangle$ including quantum statistics for the toy model introduced in eq.~(\ref{eq:toy}), 
as a function of the 
temperature and for different DM masses. To avoid a discontinuity due to the abrupt change between the two different prescriptions for 
$\sqrt{s} < 2 \, \mathrm{GeV}$ above and below the QCD phase transition, we make an interpolation of the form $\langle \sigma v \rangle = a T + b$ across the grey shaded band. 
For comparison, we also indicate (with dot-dashed lines) the thermally 
averaged cross sections that one would obtain when ignoring the QCD phase transition, i.e.\ when considering 
annihilation into free quarks and gluons even at low temperatures. 
We find that doing so significantly overestimates the DM production rate for small DM masses and small temperatures.
In particular, it is clear that 
hadronic decays are kinematically forbidden for $\sqrt{s} < 2 m_\pi$, leading to a substantial suppression of 
$\langle\sigma v\rangle$ at small temperatures compared to the naive estimate based on free light quarks. Indeed, for 
$m_e < m_S < m_\mu$ and in the limit $T \to 0$, only annihilations into electrons give a relevant contribution to the 
thermally averaged annihilation cross section, such that $\langle \sigma v \rangle$ becomes almost independent of 
$m_S$ in this parameter region. For the largest values of $m_S$ considered in the right panel of 
figure~\ref{fig:gammah_updated} on the other hand, the CMS energy remains large enough even for small 
temperatures that we can consider annihilations into free quarks and gluons. As a result, including the QCD phase transition makes almost no difference and the annihilation cross section remains large even for $T \to 0$.

We finally note that for the smallest DM masses considered in the right panel of figure~\ref{fig:gammah_updated}, the thermally averaged annihilation cross section exhibits a minimum around $T \approx 10\,\mathrm{MeV}$ and then rises again slightly towards smaller temperatures (most clearly visible for $m_s = 10\,\mathrm{MeV}$). This is a result of the DM particles still being semi-relativistic at these temperatures, such that $v_\text{lab}$ is 
(by up to a factor of 2) smaller than the CMS relative velocity $v_\text{cms}$, while the two 
velocities agree for smaller temperatures, i.e.\ in the non-relativistic limit. 
Note that for $m_s = 1\,\mathrm{MeV}$ the same effect is present but is partially compensated by the phase space suppression for annihilations into electrons at small temperatures.

\section{Freeze-in of scalar singlet dark matter}
\label{sec:scalarsinglet}

We now apply the largely model-independent formalism outlined in the previous sections to a specific 
DM model. For this purpose we consider a new real singlet scalar $S$, which is stabilised by a 
$\mathbb{Z}_2$ symmetry. The most general renormalisable Lagrangian is then
\begin{equation}
\mathcal{L} = \frac{1}{2} \partial_\mu S \partial^\mu S + \frac{1}{2} \mu_{S}^2 S^2 + \frac{1}{2} \lambda_{hs} S^2|H|^2 + \frac{1}{4} \lambda_{s} S^4 \, .
\end{equation}
After EWSB the term involving the Higgs field induces terms proportional to $h^2 S^2$, $v hS^2$ 
and $v S^2$.  The latter gives a contribution to the scalar singlet mass, which as a result is given by
\begin{equation}
\label{m_S_tree}
m_S(T) = \sqrt{\mu_{S}^2 + \frac{1}{2}{\lambda_{hs} v(T)^2}} \, .
\end{equation}
This effect leads to a temperature dependence of the mass term even if the scalar singlet is not in 
equilibrium with the SM thermal bath. 

In the following we will be interested in the case where the phenomenology of the model is driven by 
$m_s$ and $\lambda_{hs}$. In particular, we assume that
$\lambda_{hs}$ is sufficiently small that the scalar singlet never 
entered into thermal equilibrium with the SM heat bath and that its relic abundance is determined by 
the freeze-in mechanism.\footnote{%
The regime where scalar singlet DM is produced via the freeze-out mechanism has been extensively 
studied elsewhere~\cite{Silveira:1985rk,McDonald:1993ex,Cline:2013gha,GAMBIT:2017gge,Binder:2017rgn}. For a general discussion of the transition between freeze-in and freeze-out, we refer to Ref.~\cite{Du:2021jcj}.
}
The latter requirement also means that the quartic self-coupling $\lambda_s$  should 
be small enough to avoid equilibration of the scalar singlet with 
itself via $2\leftrightarrow4$ processes~\cite{Carlson:1992fn,Bernal:2018ins}. Assuming that the scalar singlets account for all of the DM in the universe, this requirement translates to the relatively weak upper bound $\lambda_s \lesssim 10 (m_s / \mathrm{GeV})$. For comparison, the typical bound on DM self-interactions, $\sigma / m_s \lesssim 1 \, \mathrm{cm^2 /g}$~\cite{Clowe:2006eq}, translates to $\lambda_s \lesssim 100 (m_s / \mathrm{GeV})^{3/2}$ for small DM masses.

The processes that contribute to the freeze-in yield are fundamentally different before and after the EWPT. In the 
former case, the only process that leads to the production of scalar singlets is $H H \to S S$, which 
in our approach is calculated by considering the annihilation cross section for $S S \to H H$. In the 
latter case, on the other hand, a 
multitude of SM states can contribute and we need to calculate the annihilation cross section for 
processes like $S S \to h^\ast \to f \bar{f}$. 
Once all annihilation rates have been calculated, we can simply integrate the right-hand side of eq.~(\ref{eq:dYdx}) over the relevant range of $x$ in order to 
obtain the final abundance $Y_s$.

For $m_S < m_h / 2$ (and sufficiently high reheating temperature) one finds that the dominant contribution to the scalar singlet yield stems from temperatures $T \sim m_h / 2$. This can equivalently be interpreted as either equilibrium decays of SM Higgs bosons or annihilations enhanced by an $s$-channel resonance (see section~\ref{sec:FI_decays}). For $m_S > m_h/2$, on the other hand, there is no such resonant enhancement, as the decays of on-shell Higgs bosons into scalar singlets are kinematically forbidden. In this case freeze-in production proceeds dominantly via off-shell Higgs decays at higher temperatures, such that the thermal effects discussed in sections~\ref{sec:finiteT} and \ref{sec:hdecay} become particularly relevant.

Another interesting scenario is when the reheating temperature $T_\text{RH}$ is small compared to the Higgs boson mass: $T_\text{RH} \ll m_h$. In this case the density of Higgs bosons in the thermal plasma is exponentially suppressed for all relevant temperatures and there is no resonant enhancement of the freeze-in production even for $m_S < m_h/2$. Instead, the processes relevant for the freeze-in production of scalar 
singlets can be written as contributing via an effective dimension-5 operator of the form
\begin{equation}
 \mathcal{L} \supset \frac{1}{\Lambda_f} \bar{f} f S^2 \, , \label{eq:dim5}
\end{equation}
where $\Lambda_f = m_h^2 / (\lambda_{hs} m_f)$. 
As a direct consequence, we will see that the freeze-in yield becomes sensitive to the reheating temperature
-- as expected whenever a non-renormalizable operator is responsible for the DM production~\cite{Hall:2009bx}. 
We will consider both of these cases in turn in the following.

\subsection{High reheating temperature}
\label{sec:highTR}

For the case that $T_\text{RH} \gg m_S, m_h$ the freeze-in production is infrared-dominated, meaning that the 
resulting abundance is independent of the reheating temperature. This follows from the observation that before 
EWSB and for $T \gg m_S,m_h$ the DM production cross section is proportional to $1/s$, such that the DM 
production rate is proportional to the temperature, $n_\chi \langle \sigma v\rangle\propto T$, and therefore 
becomes negligible compared to the Hubble expansion rate at high enough temperatures.

\begin{figure}[t]
  \centering 
  \includegraphics[width=0.49\textwidth]{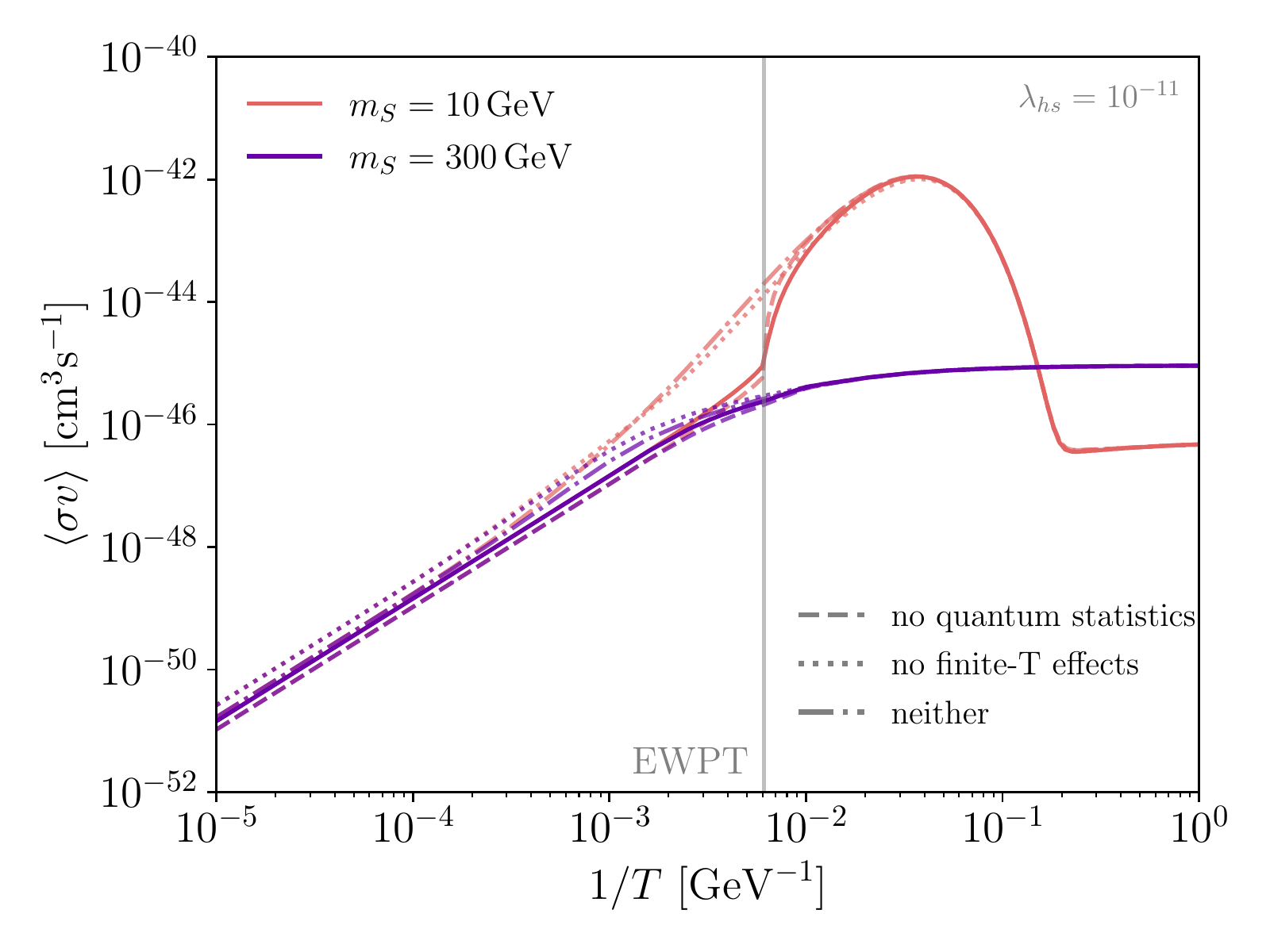}
  \includegraphics[width=0.49\textwidth]{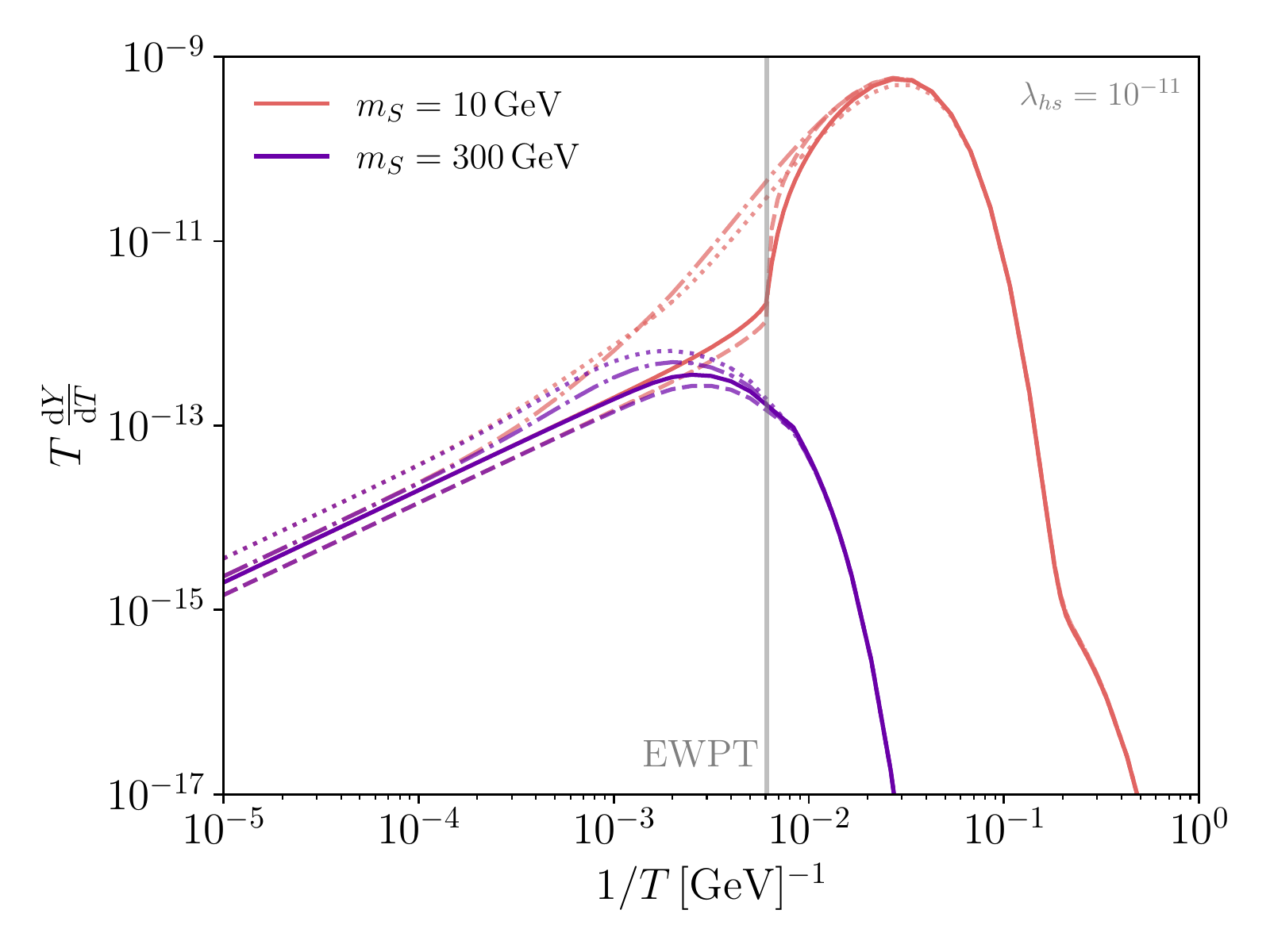}
  \caption{Thermally averaged annihilation cross section (left) and the rate of change of the scalar singlet yield 
  $T dY/dT$ (right) for $\lambda_{hs} = 10^{-11}$ and $m_S=10$ GeV (red) and $m_S=300$ GeV (purple). 
  Dashed, dotted and dash-dotted lines correspond respectively to the cases where we neglect quantum 
  statistics, other thermal effects, or both. The grey vertical line indicates $T=T_{\rm EW}$. }
  \label{fig:rates}
\end{figure}

In figure~\ref{fig:rates} we show the thermally averaged DM annihilation cross section (left) and the resulting 
change in the DM yield $dY/dx = T dY/dT$ (right) as a function of inverse temperature for two representative
DM masses below ($m_S = 10\, \mathrm{GeV}$) and above ($m_S = 300 \, \mathrm{GeV}$) the Higgs
resonance, respectively. To highlight the importance of in-medium effects, we also show the results that one 
would obtain when neglecting quantum statistics (dashed lines), when neglecting thermal effects (dotted) and 
when neglecting both (dot-dashed lines). 

As expected we find qualitatively different behaviour for the two DM masses. For $m_S = 10 \, \mathrm{GeV}$ production is dominated by processes involving the exchange of an on-shell Higgs boson. Hence, the thermally averaged cross section and the production rate receive a strong enhancement 
when the typical CMS energy in the thermal bath is comparable to the Higgs boson mass. For $m_S = 300 \, \mathrm{GeV}$, on the other hand, the virtual Higgs boson must always be off-shell and hence the temperature dependence of the annihilation cross section becomes more trivial: For large temperatures the cross section falls proportional to $1/s \propto 1/T^2$, while for small temperatures it becomes constant
as expected for $s$-wave annihilation. In both cases the corresponding production rate exhibits an exponential suppression when the temperature drops below the DM mass because of the additional factor of $(n^{\rm MB})^2$ in eq.~\eqref{eq:dYdx}. We note that for $m_S = 300 \, \mathrm{GeV}$ relevant contributions to the DM abundance arise from both before and after the EWPT (indicated by the vertical line).

We find that the inclusion of quantum statistics leads to a visible enhancement of the annihilation cross section at high temperatures ($T \gtrsim m_h$), when all relevant initial and final states are bosonic. For smaller temperatures, on the other hand, there are two competing effects: a suppression arising from the fermionic nature of the quarks and leptons in the final state and an enhancement arising from the bosonic nature of the $s$-channel resonance (see eq.~\eqref{eq:width_prescription}). These two effects cancel approximately, leading to only a small net impact from including quantum statistics. Thermal effects are particularly important above the electroweak phase transition, where there is no longer a resonant enhancement for $m_S = 10\,\mathrm{GeV}$. Moreover, the thermal mass of the complex Higgs field has a relevant effect by reducing the available phase space for the annihilation process.

\begin{figure}[t]
  \centering 
  \includegraphics[width=0.49\textwidth]{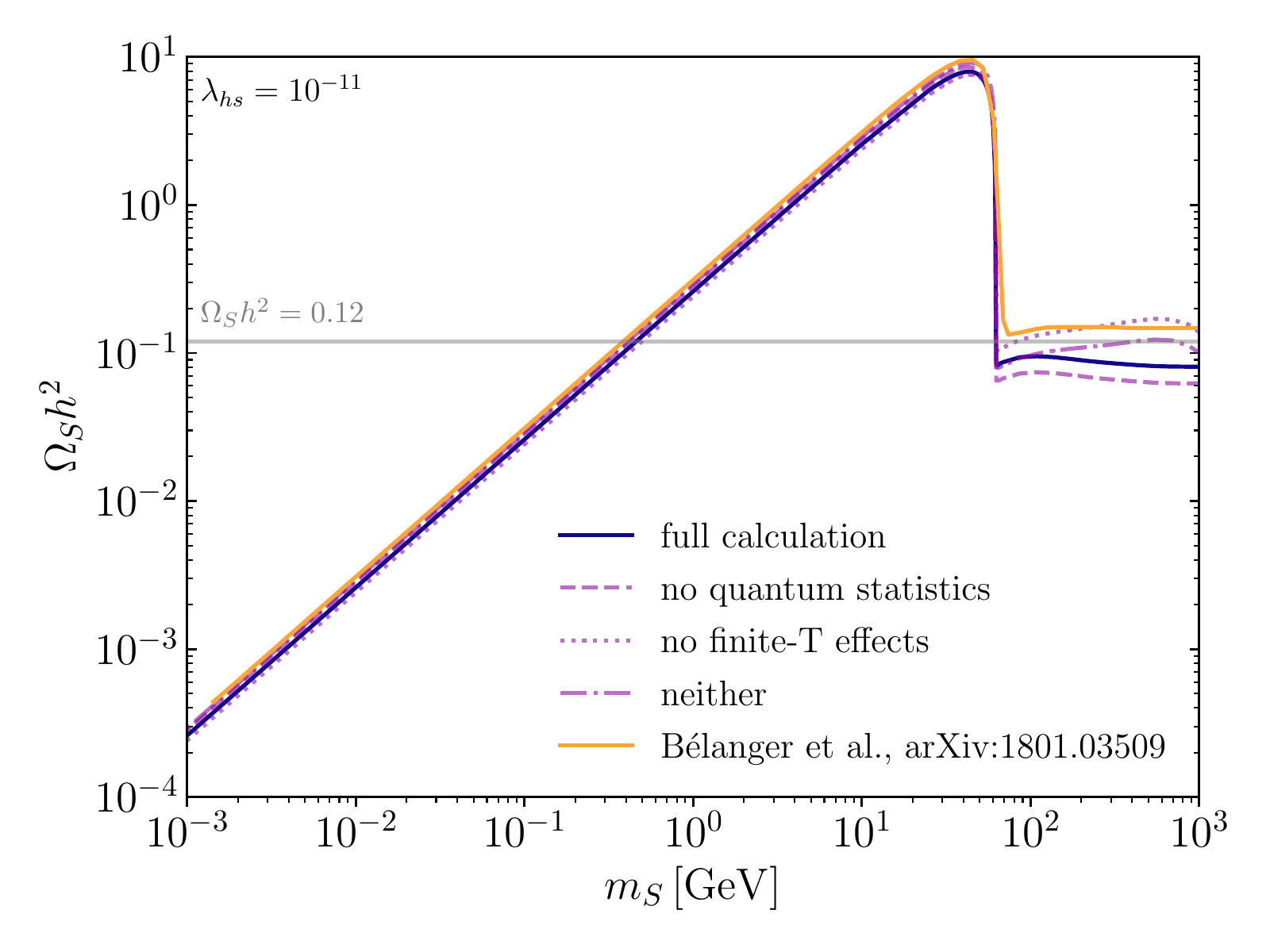}
  \includegraphics[width=0.49\textwidth]{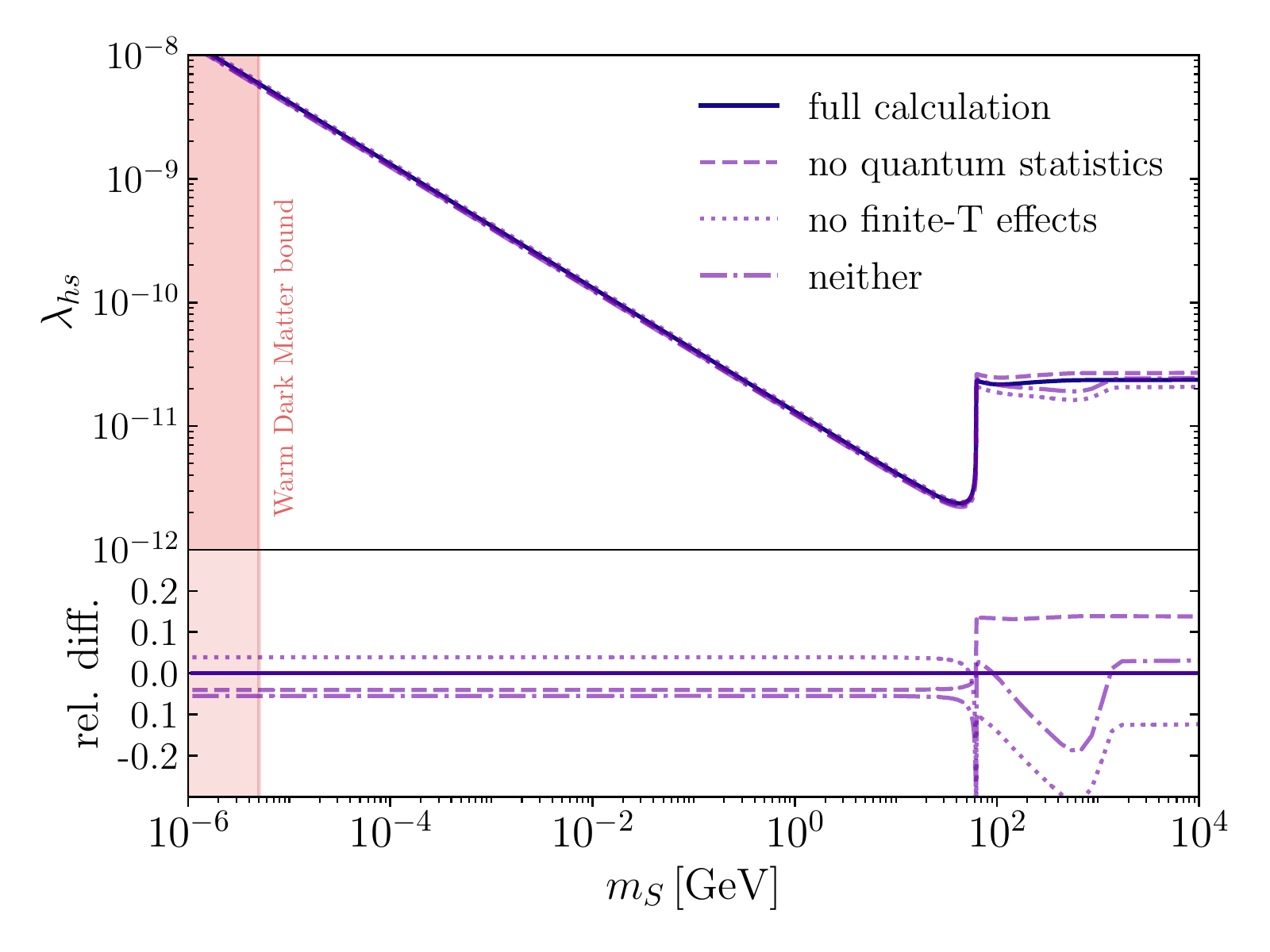}
  \caption{{\it Left:} Relic density $\Omega h^2$ as a function of the scalar singlet mass $m_S$ for a fixed value of 
  the portal coupling $\lambda_{hs}=10^{-11}$ in the case of a {\it high reheating} temperature. We compare the results obtained from \ds\ (purple lines) with the ones obtained in ref.~\cite{Belanger:2018ccd}, which includes quantum statistics using micrOMEGAs but no (other) thermal effects. {\it Right:} Portal coupling $\lambda_{hs}$ needed to reproduce the observed DM relic abundance $\Omega_S h^2 = 0.12$ as a function of $m_S$. The smaller panel in the bottom shows the relative difference when the various effects considered in this work are switched off.}
   \label{fig:relic}
\end{figure}

By integrating the curves shown in the right panel of figure~\ref{fig:rates} we can calculate the DM relic abundance $\Omega_S$ for given values of $\lambda_{hs}$ and $m_S$. We show in the left panel of figure~\ref{fig:relic} the result for $\lambda_{hs} = 10^{-11}$. For comparison, we again show the various curves without in-medium effects and the one obtained in ref.~\cite{Belanger:2018ccd} using micrOMEGAs  
when including quantum statistics (but neglecting thermal effects). We find that for $m_S < m_h/2$ the different curves are quite close to each other, but that there is a small difference between our result without finite temperature effects (dotted line) 
and that obtained in ref.~\cite{Belanger:2018ccd}. This difference can be traced back to the slightly different treatment of the $s$-channel resonance as well as the updated Higgs mass value $m_h= 125.25\,\mathrm{GeV}$~\cite{ParticleDataGroup:2020ssz}. For higher masses, the difference between the various curves is more pronounced, with thermal effects having a particularly sizable effect on the predicted relic abundance.

We conclude our discussion of the high reheating temperature case by showing, in the right panel of 
figure~\ref{fig:relic}, the value of $\lambda_{hs}$ that is needed to reproduce the observed DM relic abundance 
$\Omega_S h^2 = 0.120 \pm 0.001$~\cite{Planck:2018vyg} as a function of $m_S$. 
We note that this exercise constitutes an almost trivial rescaling of the left panel of the figure because
$t$- and $u$-channel diagrams contributing to $SS\to hh$ are highly suppressed for such small values of 
$\lambda_{hs}$; we therefore find that $\sigma v\propto \lambda_{hs}^2$, and hence 
$\Omega_S h^2\propto \lambda_{hs}^{-1/2}$, to an excellent accuracy. We also indicate, with the
same line style as before, the individual impact of the various finite-temperature effects that we have implemented here;
the smaller plot at the bottom shows the relative difference compared to the full treatment (solid purple line). 
We find that these differences can be as large as $30\%$  for $m_S > m_h/2$, corresponding to a difference
in $\Omega_S h^2$ of almost a factor of two; for smaller masses, the impact on the relic density remains
below 10\%. 
Finally, while the freeze-in mechanism in principle works for arbitrarily small singlet masses, the resulting particles 
at some point become too relativistic to be consistent with structure formation and the observed matter power 
spectrum. In a recent analysis 
of scalar singlet DM the resulting bound on $m_s$, essentially a warm DM bound, 
was found to be $m_S > 4.4 \, \mathrm{keV}$ 
($m_S > 5.3\,\mathrm{keV}$) if $\lambda_s$ is large enough (small enough) that DM particles experience (do not 
experience) self-interactions~\cite{Egana-Ugrinovic:2021gnu} (see also Ref.~\cite{Yunis:2021kgo}). We indicate this bound by the red shaded region, 
noting that the warm DM bound for FIMPs is generally a bit more stringent than for DM produced via 
freeze-out~\cite{DEramo:2020gpr,Dvorkin:2020xga,Decant:2021mhj}.

\subsection{Low reheating temperature}
\label{sec:lowTR}

In the discussion so far we have assumed that the reheating temperature is large enough that it becomes 
irrelevant for the freeze-in calculation. However, there is strictly speaking no observational evidence for such 
large reheating temperatures, which may be as low as $5\,\mathrm{MeV}$ without conflict with data~\cite{deSalas:2015glj}. For $T<T_\text{RH} \ll m_h$ the interactions between scalar singlets and SM fermions are described by the effective dimension-5 operator given in eq.~\eqref{eq:dim5}.\footnote{As pointed out recently~\cite{Frangipane:2021rtf}, it is not sufficient to require $T_\text{RH} < m_h$ for the EFT description to be valid, since particles in the tail of the Boltzmann distribution can still experience resonant enhancement. From the left panel of figure~\ref{fig:rates} we can infer that the resonance ceases to be relevant for $T \lesssim 5 \, \mathrm{GeV}$.}  As a result, the annihilation cross section scales proportional to $\Lambda_f^{-2}$ and does not depend on temperature as long as $m_S, m_f < T$. The DM production rate then scales proportional to $T^3$ and therefore drops faster than the Hubble rate. This implies that freeze-in production is ultraviolet-dominated, i.e.\ the DM relic abundance will be directly sensitive to the reheating temperature.
 
With decreasing temperature fewer and fewer SM final states $f$ will be kinematically accessible, leading to an exponential suppression of the annihilation cross section for $T < m_f$.  Nevertheless, since scalar singlets couple to SM fermions proportional to their mass, these final states may still give a relevant contribution to the freeze-in yield. 
In this case the production rate decreases even more rapidly than $T^3$, i.e.\ production will be even more strongly peaked in the UV. In this context, it is essential to account for the formation of hadronic bound states after the QCD phase transition in order to obtain realistic estimates of the off-shell Higgs decay width. Having 
implemented all of these effects, as detailed in section \ref {sec:hdecay}, we are now in the position to extend our 
discussion of the freeze-in production of scalar singlets also to the case of low reheating temperatures.

\begin{figure}
  \centering
  \includegraphics[width=0.49\textwidth]{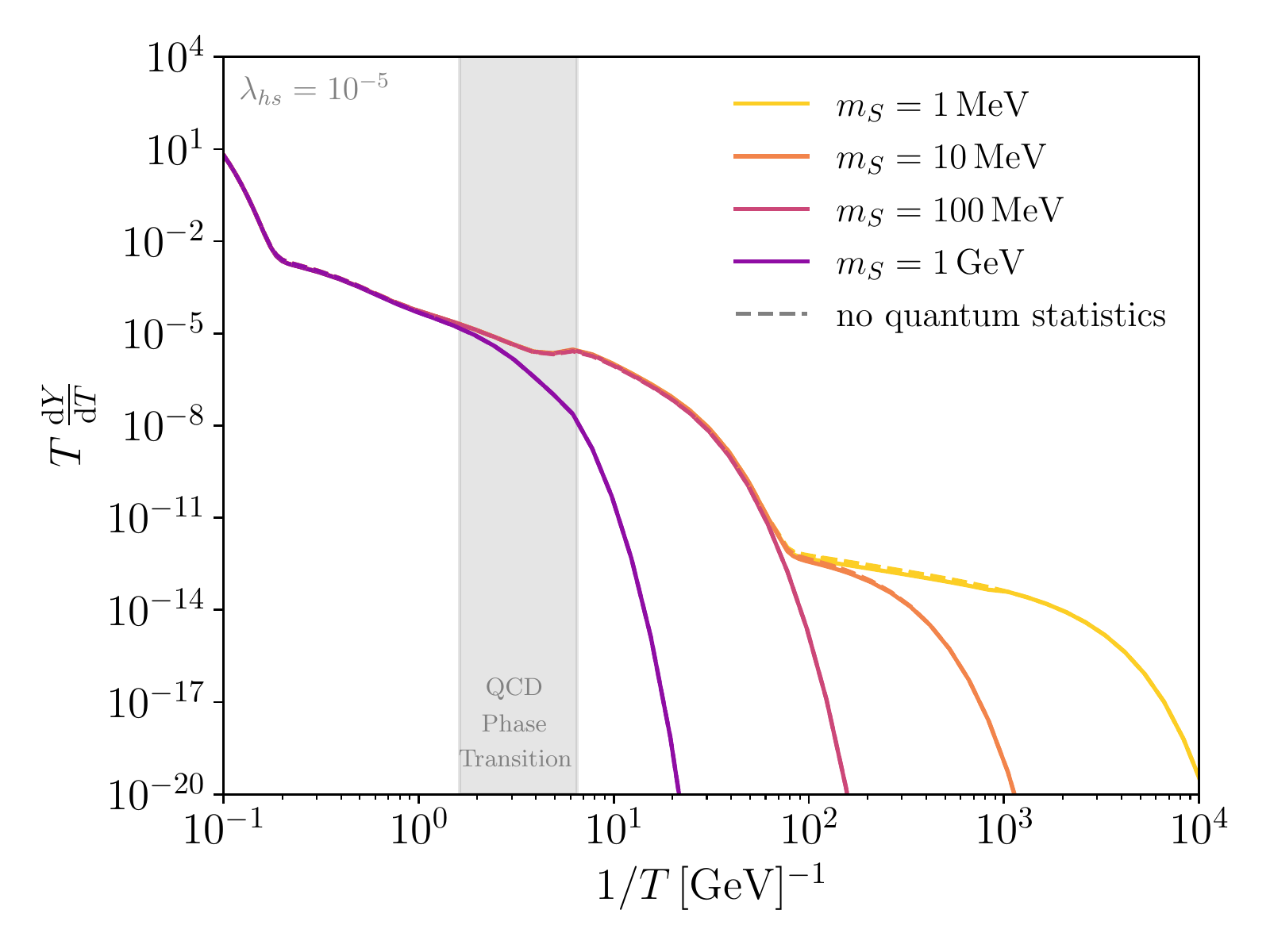}
  \includegraphics[width=0.49\textwidth]{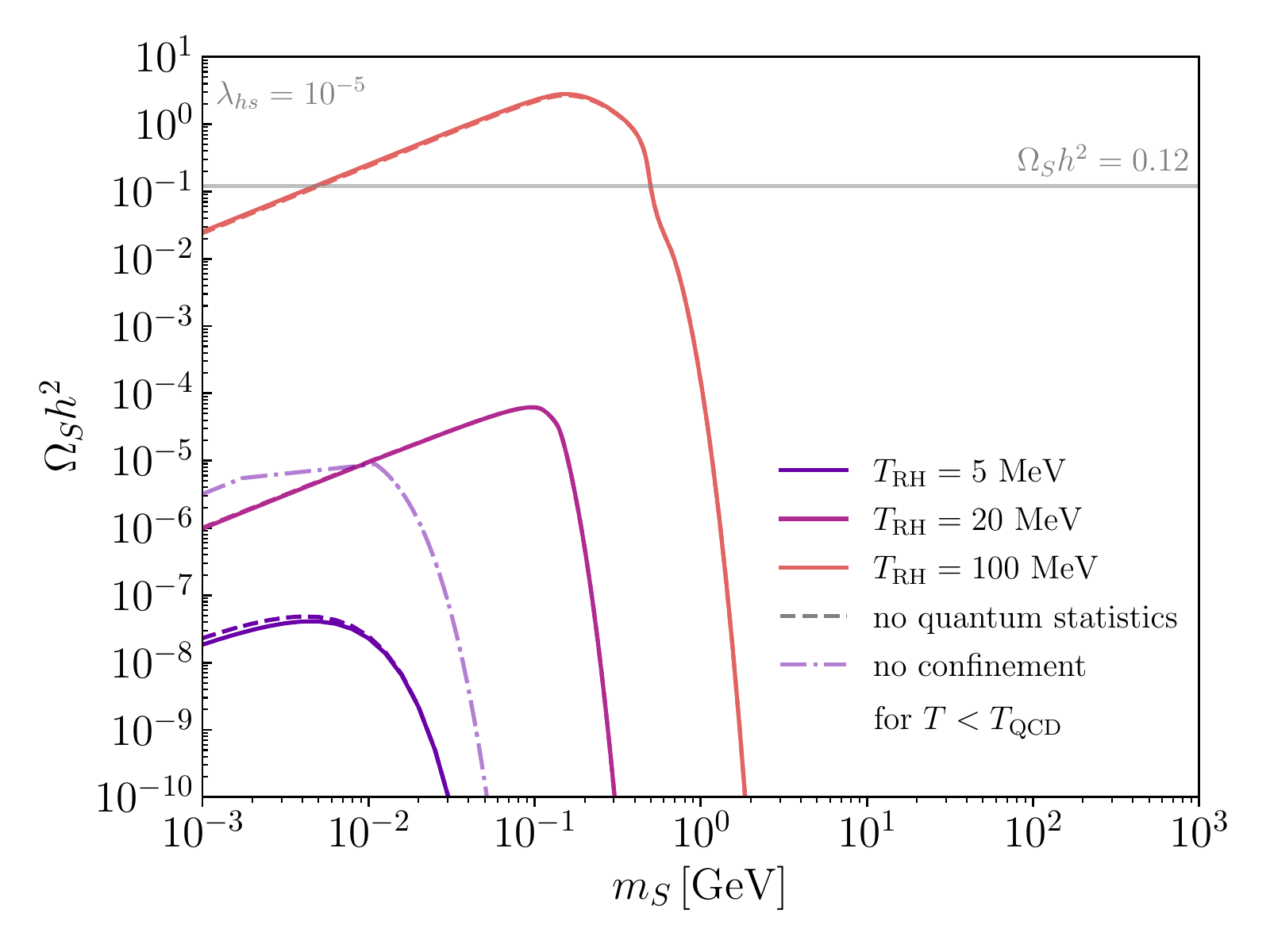}
  \caption{
  {\it Left:} Rate of change of the scalar singlet yield as a function of inverse temperature for 
  $\lambda_{hs} = 10^{-5}$ 
  and different values of the singlet mass as indicated. The description as an effective dimension-5 operator
  is valid for $T\gtrsim5$\,GeV,  while the increase of the rate for smaller temperatures is caused by the 
  resonance (cf.~the right panel of figure \ref{fig:rates}).
  {\it Right:} Resulting relic density when integrating the differential yield over $T < T_\text{RH}$ as a function of the singlet mass and for different values of the reheating temperature. In both panels dashed lines indicate the (small) effect from neglecting quantum statistics.
  For comparison, the dash-dotted line ignores confinement, but includes quantum statistics effects.
  }
  \label{fig:relic_lowTR}
\end{figure}

In the left panel of figure~\ref{fig:relic_lowTR} we show the differential yield $T dY/dT$ as a function of inverse temperature for different values of $m_S$ and fixed coupling $\lambda_{hs} = 10^{-5}$,
effectively extending the right panel of figure \ref{fig:rates} to smaller temperatures (note the change in slope
at $T\sim5$\,GeV in both figures, indicating the onset of the regime where the effective operator description
is valid).
 These lines are obtained by substituting the thermally averaged annihilation cross  shown in the right panel of figure~\ref{fig:gammah_updated} into eq.~(\ref{eq:dYdx}). As before, this leads to an exponential suppression of the differential yield for $T < m_S$. Conversely, for sufficiently large temperatures ($T > m_S$) the production rate becomes independent of the DM mass. As anticipated, the production rate drops steeply with decreasing temperature, such that the dominant contribution to DM production comes from $T \approx T_\mathrm{RH}$.

For small DM masses we therefore find that $Y_s$ is approximately independent of $m_S$ and hence $\Omega_S h^2 \propto m_S$. This is confirmed in the right panel of figure~\ref{fig:relic_lowTR}, which shows $\Omega h^2$ as a function of $m_S$ for fixed coupling $\lambda_{hs} = 10^{-5}$ and different values of $T_\text{RH}$. For $m_S > T_\text{RH}$, on the other hand, the production becomes exponentially suppressed and the resulting relic abundance drops sharply. For fixed $m_S$ the relic abundance depends monotonically on $T_\text{RH}$. For values of $T_\text{RH}$ close to the lower bound of $5 \, \mathrm{MeV}$ the resulting abundance is very small unless $\lambda_{hs}$ is increased significantly. We emphasise that in the case of small reheating temperature, the DM production rate is dominated by the heaviest state kinematically accessible, which are typically non-relativistic. The effect of neglecting quantum statistics (dashed lines) is therefore considerably 
smaller than for the case of high reheating temperature considered above, cf.~figure \ref{fig:relic}. However, a proper treatment of the QCD phase transition is essential, as illustrated by the dash-dotted line 
(shown for clarity only for one of the parameter choices).

\begin{figure}
\centering
\includegraphics[width=0.49\textwidth]{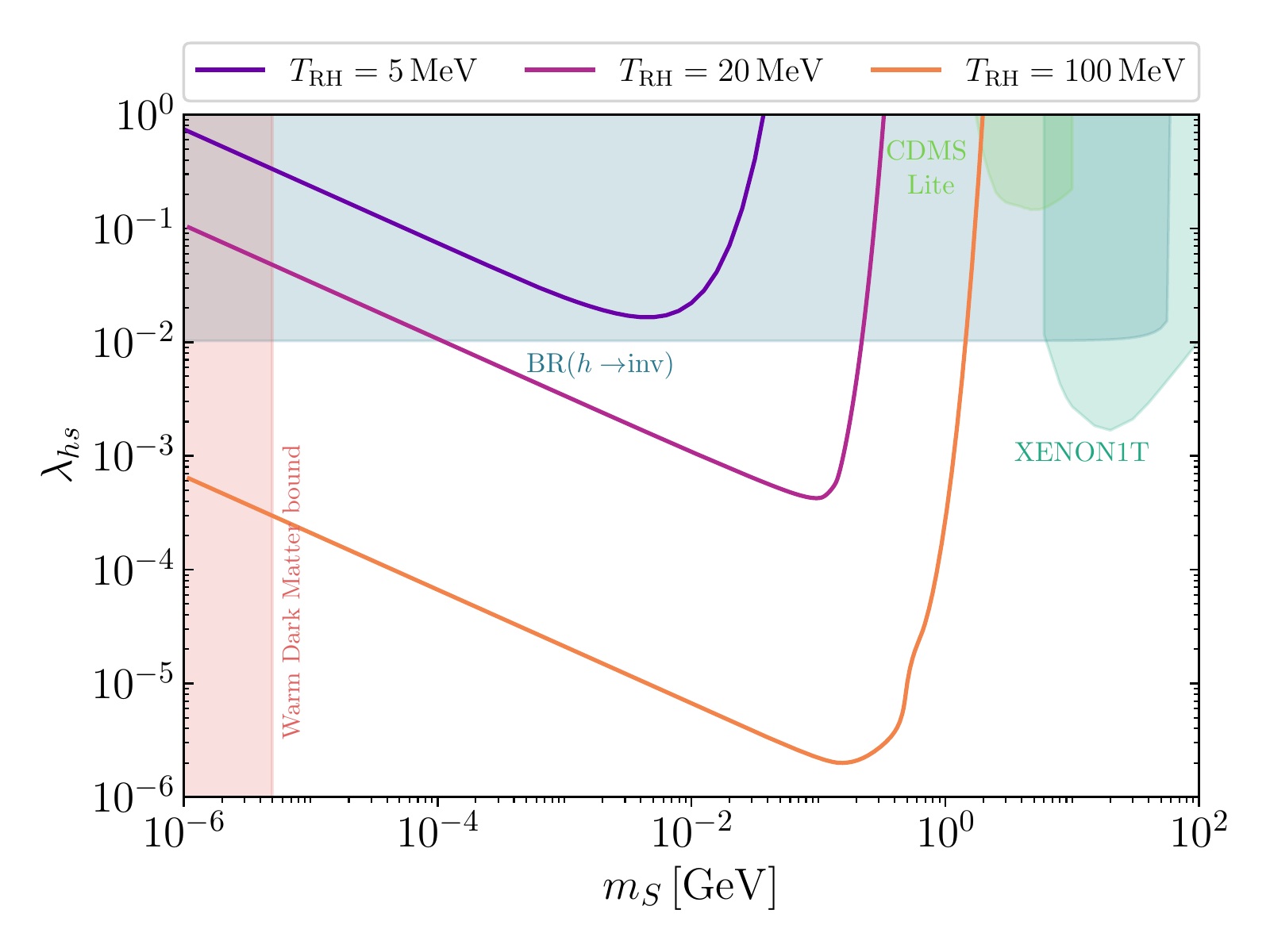}
\includegraphics[width=0.49\textwidth]{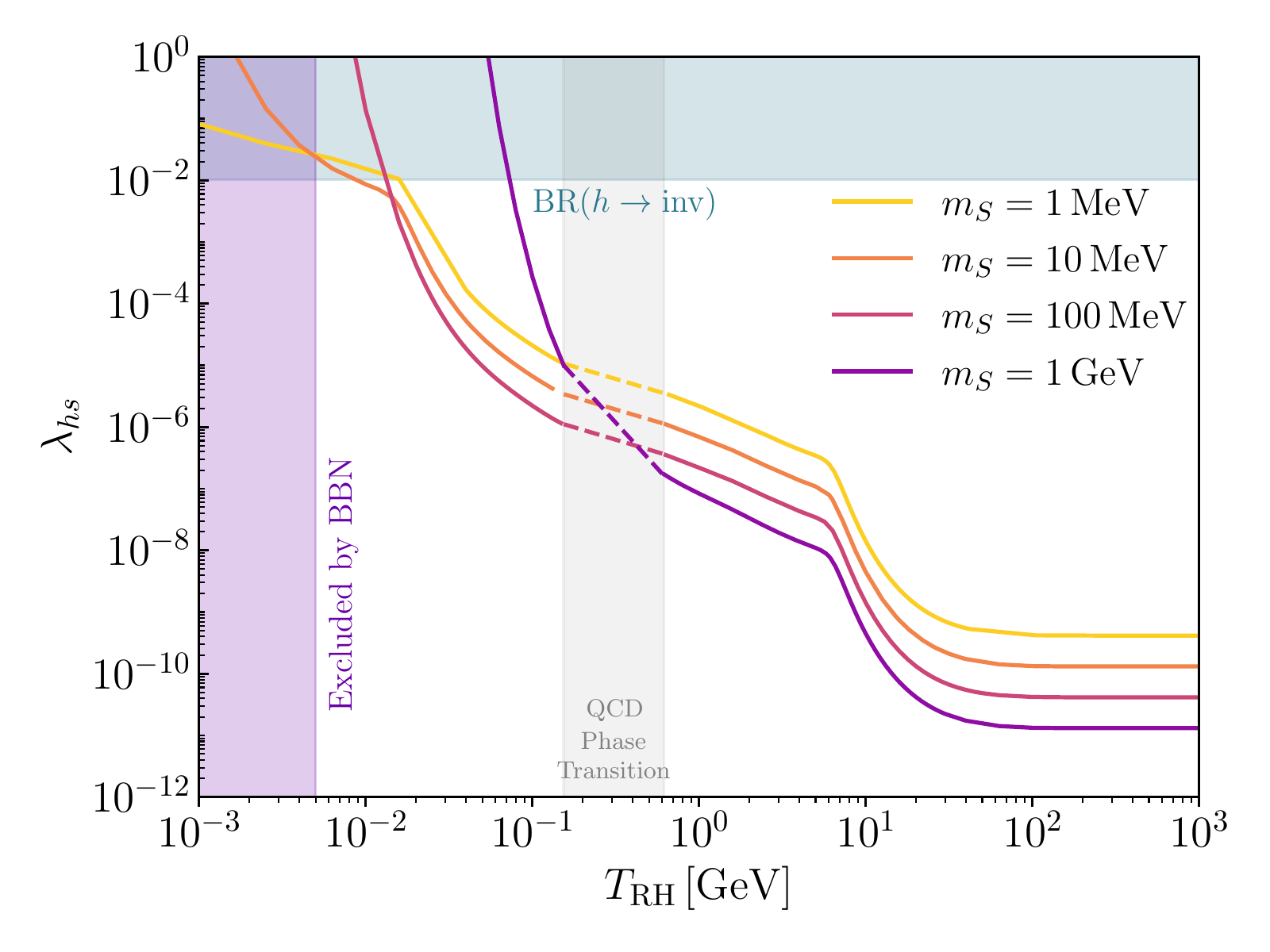}
\caption{\label{fig:lambda_lowTR}\textit{Left:} Value of the portal coupling $\lambda_{hs}$ that leads to the observed DM relic 
abundance as a function of the singlet mass $m_S$ and for different {\it low reheating} temperatures 
$T_\text{RH}$. In 
addition we show bounds from direct detection experiments~\cite{SuperCDMS:2013eoh,XENON:2018voc}, 
the LHC~\cite{ATLAS:2020kdi} and the Lyman-$\alpha$ 
forest~\cite{Egana-Ugrinovic:2021gnu,DEramo:2020gpr}. See text for 
details. \textit{Right:} Value of the portal coupling plotted as a function of the reheating temperature, $T_\mathrm{RH}$, for different singlet masses. $T_\mathrm{RH}<5\,\mathrm{MeV}$ (shown in purple) is excluded by BBN~\cite{deSalas:2015glj}. The grey band corresponds to the temperature range we consider for the QCD phase transition. See text for details.
}
\end{figure}

In figure~\ref{fig:lambda_lowTR} we finally show the coupling required to reproduce the observed DM relic abundance as a function of $m_S$ for different values of $T_\text{RH}$. As expected, we find that for the smallest values of the reheating temperature that we consider ($T_\text{RH} < 20 \, \mathrm{MeV}$) the portal coupling must be rather large in order to reproduce the observed DM relic abundance ($\lambda_{hs} \gtrsim 10^{-2}$). At first sight this leads to an apparent inconsistency, given that such large portal couplings should cause the dark sector to thermalise with the SM. However, this is only true for large reheating temperatures. For small temperatures the interaction rate is suppressed proportional to $T^2 v^2 / m_h^4 \ll 1$ such that one can consistently apply the freeze-in formalism even for rather large portal couplings.

Nevertheless, for sufficiently large portal couplings it becomes feasible to search for scalar singlets in the laboratory, in particular at the LHC where the CMS energy is sufficient to produce on-shell Higgs bosons and hence the cross section for DM production may be much larger than in the early universe. The most promising strategy to probe sub-GeV scalar singlets are therefore searches for invisible Higgs decays. The corresponding partial decay width is given by
\begin{equation}
 \Gamma(h \to SS) = \frac{\lambda_{hs}^2 v^2}{32\pi m_h}\sqrt{1- \frac{4m_S^2}{m_h^2}} \, ,
\end{equation}
from which the invisible branching ratio can be calculated as
\begin{equation}
\text{BR}(h \to \text{inv}) = \frac{ \Gamma(h \to SS)}{ \Gamma(h \to SS) + \Gamma_{h,\text{SM}}} \,. 
\end{equation}
A recent combination of searches for invisible Higgs decays at the LHC yields $\text{BR}(h \to \text{inv}) < 0.11$~\cite{ATLAS:2020kdi}, which translates to $\lambda_{hs} < 0.01$. In other words, for the smallest values of $T_\text{RH}$ considered in our analysis, the coupling required by the freeze-in mechanism is already excluded by LHC data.

For singlet masses at the GeV scale there are also relevant constraints from direct detection experiments, which probe the spin-independent scattering cross section
\begin{equation}
 \sigma_N = \frac{m_N^4}{4\pi (m_S + m_N)^2} \frac{\lambda_{hs}^2 f_N^2}{m_h^4} \, 
\end{equation}
with $f_N \approx 0.3$ being the effective Higgs-nucleon coupling~\cite{Cline:2013gha} and $m_N$ denoting the nucleon mass. Figure~\ref{fig:lambda_lowTR} shows the exclusion limits from some of the most constraining experiments~\cite{SuperCDMS:2013eoh,XENON:2018voc}. We find these exclusions to be irrelevant unless $m_S \gg T_\text{RH}$ and the freeze-in production of scalar singlets is exponentially suppressed. Finally, we show once again the warm DM constraints from ref.~\cite{Egana-Ugrinovic:2021gnu}, which place a lower bound on the allowed range of scalar singlet masses.

Finally, in the right panel of figure~\ref{fig:lambda_lowTR} we show instead the dependence on $\lambda_{hs}$ on the reheating temperature $T_\mathrm{RH}$ for several different values of $m_S$. As expected, for large enough values of $T_\mathrm{RH}$ the predicted abundance (and hence the required value of $\lambda_{hs}$) becomes independent of $T_\mathrm{RH}$. We find that this is the case for $T_\mathrm{RH} \gtrsim 50 \, \mathrm{GeV}$, i.e.\ once the reheating temperature is greater than about half the Higgs boson mass. For smaller values of $T_\mathrm{RH}$ the required value of $\lambda_{hs}$ increases rapidly and becomes incompatible with LHC bounds on the Higgs boson invisible width for small $T_\mathrm{RH}$, with the precise bound depending on $m_S$. The lower bound $T_\mathrm{RH} \gtrsim 5 \, \mathrm{MeV}$~\cite{deSalas:2015glj} from Big Bang Nucleosynthesis (BBN) is indicated by the violet exclusion region.

\section{Conclusions}
\label{sec:conclusions}
Over the last few years the topic of DM relic density calculations has diversified significantly, with many new ideas 
being discussed that go beyond the conventional freeze-out mechanism. At the same time, the topic has matured 
in the sense that order-of-magnitude estimates are no longer sufficient and sub-leading effects need to be studied 
and included. Publicly available software tools for automatic relic density calculations are challenged to keep up 
with these rapid developments and the increasing complexity of calculations. At the same time, the need to solve 
Boltzmann equations efficiently and accurately is greater than ever. In the present work we have addressed this 
need by considering in detail a number of effects relevant for the freeze-in mechanism and included them in the 
latest release of \ds.

In the first part of our study we performed a general analysis of the freeze-in formalism, showing that it is possible 
to express the DM production rate relevant for the freeze-in mechanism in terms of the DM annihilation rate even 
when including quantum statistics and other temperature-dependent effects. 
This makes it possible to write the collision operator in a simple form that 
allows for an efficient and modular numerical implementation. We extended our discussion to the case of freeze-
in production from decays and argued that the resulting expressions are equivalent to the case of an $s$-channel 
resonance in the DM annihilation cross section. Well-established techniques for calculating the latter can 
therefore be directly applied also to freeze-in calculations.

A particularly interesting example are DM models that couple to the SM via the Higgs field. In this case we can 
make use of precision calculations of the off-shell Higgs decay width to obtain the DM annihilation cross section. 
To account for the wide range of temperatures relevant for freeze-in, we have discussed in detail how finite 
temperatures modify the Higgs decay widths, with a special focus on the electroweak and QCD phase transitions. 
Moreover, we propose a new way to include higher-order corrections for large centre-of-mass energies that 
avoids the issue of unitarity violation.

We have applied our improved freeze-in formalism to the case of scalar singlet DM. For large 
reheating temperatures, freeze-in production of scalar singlets has been discussed in great detail in the literature, 
but we improve upon these earlier studies in terms of the speed and precision of our calculations and the range of 
effects considered. In particular, we show that for scalar singlet masses above the Higgs resonance it is important 
to include an accurate treatment of the electroweak phase transition.

Finally, we have considered an alternative freeze-in scenario, in which the reheating temperature is small 
compared to the Higgs boson mass and freeze-in production proceeds via non-renormalisable effective 
operators. In this case the DM relic abundance is sensitive to the reheating temperature and a consistent 
implementation of the QCD phase transition is essential. Intriguingly, we find that for the smallest reheating 
temperatures considered, the couplings implied by the observed DM relic abundance can be probed by LHC 
measurements of the SM Higgs boson invisible branching ratio.

Together with this study we release a new version 6.3~of \ds, where the various in-medium effects and our improved 
calculation of the Higgs decay widths have been implemented. The modularity of the code makes it possible to 
apply our findings to different freeze-in scenarios, for which fast and accurate relic density calculations are 
desired.\footnote{%
For reference, the curves in figure \ref{fig:relic}, which are based on 52 grid points, were obtained in $752$\,s (full calculation), $1828$\,s (only including quantum statistics), $36$\,s (only including thermal effects) and $32$\,s (ignoring all medium effects). The program to calculate these curves, {\code examples/aux/FreezeIn\_ScalarSinglet.f}, is provided with the release as an explicit example of how to use the new \ds\ routines in practice.
} 
In future studies we plan to extend our implementation to cases where inverse processes and/or 
interactions within the dark sector cannot be neglected and the relic density calculation requires solving a set of 
coupled Boltzmann equations.
These developments will pave the way towards a unified treatment of relic density 
calculations across a wide range of different scenarios.

\acknowledgments
We thank Mikko Laine, Oleg Lebedev and Katelin Schutz for discussions and Martin W.~Winkler for providing the data from ref.~\cite{Winkler:2018qyg}. This  work  is  funded  by  the  Deutsche 
Forschungsgemeinschaft (DFG) through the Emmy Noether Grant No.\ KA 4662/1-1. TB warmly 
thanks the Albert Einstein Institute for support and hospitality during the preparation of this 
manuscript. SH acknowledges the support of the Natural Sciences and Engineering Research Council of Canada (NSERC), SAPIN-2021-00034.

\begin{appendix}

\section{DarkSUSY implementation}
\label{app:ds}

In this appendix we briefly describe the implementation of the general freeze-in 
routines in \ds\ 6.3, based on the analysis presented in this work. In order to do so let 
us first recall the case of freeze-out calculations, where the so-called invariant rate 
$W_{\rm eff}$ plays a central role in formulating the Boltzmann 
equation~\cite{Edsjo:1997bg,Gondolo:2004sc}. 
In particular, this quantity facilitates the calculation of thermal averages by splitting
the integrand into a model-independent thermal kernel and a temperature-independent part 
that  can be pre-computed to high precision, even in the presence of co-annihilations:
\be
\langle \sigma v\rangle=\int_1^\infty\!\!\! d\tilde s\,
 \frac{x\sqrt{\tilde s-1}\, K_1\!\left({2{\sqrt{\tilde s}} x}\right)}
 {2m_\chi^2  {K_2}^2(x)}
  W_{\rm eff}(s)\,.\label{eq:Weff}
\ee
At the implementation level, this is reflected by the fact that the actual relic density 
routines in \ds\ are fully model-independent, including sophisticated routines to 
tabulate $W_{\rm eff}$ for better numerical performance, while $W_{\rm eff}$ itself is provided
by the respective particle module as an interface function \code{dsanwx}.

Following the spirit of section \ref{sec:freeze-in}, and in order to make use of
existing routines for thermal averages, we thus base our implementation of the new freeze-in 
routines on a formulation that follows the freeze-out description as closely as possible.
In particular, we note that the expression for $\langle \sigma v\rangle_{\chi\chi\to\psi\psi}$
given in eq.~(\ref{eq:svav_def_full}) can be brought into the same form as eq.~(\ref{eq:Weff}) 
by introducing
\be
  W_{\rm eff}(s,T)\equiv
  16m_\chi^2 \frac{x\tilde s \sqrt{\tilde s-1}}{K_1\!\left({2{\sqrt{\tilde s}} x}\right)}
   \int_1^\infty d\gamma\, \sqrt{\gamma^2-1}e^{-2\sqrt{\tilde{s}} x \gamma}
    \sum_{\psi_1\psi_2}
   \sigma_{\chi\chi\to\psi_1\psi_2} (s,\gamma)\,.
\ee
We introduce this quantity as a new interface function \code{dsanwx\_finiteT}.
As a result, the newly implemented model-independent freeze-in routines (residing in 
\code{src/fi/}) can compute the freeze-in abundance for any particle module that provides a 
function \code{dsanwx\_finiteT}. In practice, the most important function is \code{dsfi2to2oh2},
which -- after a model has been initialised with the usual calls
to \code{dsgivemodel\_[...]} and \code{dsmodelsetup} -- returns the present DM density
resulting from the direct integration of eq.~(\ref{eq:dYdx}).\footnote{%
As advocated above, our aim is to mostly rely on already existing 
routines in \code{src/rd/} to perform the required thermal average. Since the rate
$W_{\rm eff}(s,T)$ for FIMPs is typically many orders of magnitude smaller than 
$W_{\rm eff}(T)$ for WIMPs, however, a straight-forward implementation of this idea 
inevitably causes numerical problems. We address this by {\it i)} passing a rescaled 
version of $W_{\rm eff}(s,T)$, \code{dsfianwx}, to the thermal average routines
and {\it ii)} improving the stability of the latter in the relativistic regime (which is 
irrelevant for WIMPs).
}
 Apart from performance flags (as 
explained in the function header), \code{dsfi2to2oh2} only takes the reheating temperature 
as input. Further global performance parameters are set in \code{dsfiinit}, handling, e.g.,
how potential discontinuities in the integrand of eq.~(\ref{eq:dYdx}) very close 
to phase transitions -- as explained in Section \ref{sec:finiteT} and \ref{sec:hdecay} -- 
are smoothed by linear interpolations. 

Included in the release of \ds\ 6.3 is an implementation of  \code{dsanwx\_finiteT} for the
Scalar Singlet (or \code{Silveira-Zee}) module that includes all effects described in detail in Section 
\ref{sec:scalarsinglet}. For this module, a simple call to \code{dsfiset\_silveira\_zee} makes it 
possible to individually
switch on and off the effects of quantum statistics and other finite-$T$ effects (as, e.g., explored in
figure~\ref{fig:relic}). Furthermore, we added the new particle module \code{generic\_fimp}
as a minimal demonstration of how the freeze-in abundance of a generic FIMP can be computed
with \ds. Concretely, for the sake of demonstration and in analogy to the \code{generic\_wimp} 
module, the DM particle is here assumed to couple to a single SM species; roughly reminiscent of 
situations familiar from the context of effective field theories, we further allow amplitudes
of the form $\left|\mathcal{M}\right|^2=c \left(s/\Lambda^2\right)^n$, where $c$ is the effective 
coupling strength in the regime where the stated scaling with the CMS energy is valid, 
and $\Lambda$ is the suppression scale. 

As a byproduct of the above implementation, we also expanded the list of generic 
standard model routines in \code{src\_models/common/sm/} that are accessible by all 
particle modules. For example, the newly added routines \code{dshvev\_finiteT}
and \code{dsmass\_finiteT} return temperature-dependent Higgs vev and SM masses, respectively,
as detailed in Section \ref{sec:finiteT}.
Furthermore, the function \code{dssmgammah} returning the off-shell Higgs decay width has been
updated to take into account hadronic decay products for sub-GeV center-of-mass energies,
cf.~figure~\ref{fig:gammah_updated}.

Further details about the implementation of the new functionalities and modules are provided
in the \ds\ manual as well as in the respective function headers.

\section{Analytic expressions for in-medium cross sections}
\label{app:technical}

In this appendix we collect useful expressions for the annihilation cross section in the CMS frame, 
taking into account plasma effects due to quantum statistics of the final states. First,
for DM annihilating to two heat bath particles, $\psi_1$ and $\psi_2$, the full cross section
is given by 
\be
\label{eq:sigfinal_full}
\sigma_{\chi\chi\to\psi_1\psi_2} (s,\gamma)=
\frac{N_\psi^{-1}}{8\pi s}\frac{|\mathbf{k}_{\rm CM}|}{\sqrt{s-4m_\chi^2}}
\int_{-1}^{1}\frac{d\cos\theta}{2}\left|\overline{\mathcal{M}}\right|^2_{\chi\chi\to\psi_1\psi_2} \!(s,\cos\theta)\, G_{\psi_1\psi_2}(\gamma, s, \cos\theta)\,,
\ee
where $N_\psi=2$ if $\psi=\overline{\psi}_1=\psi_2$ and  $N_\psi=1$ otherwise, and
\bea
G_{\psi_1\psi_2}(\gamma, s, \cos\theta)&=&
1+\varepsilon_\psi^2e^{-2\sqrt{\tilde{s}} x \gamma}\\
&&-\varepsilon_\psi \left\{e^{-\frac{1}{T} \left(E_{\psi_1}\gamma+|\mathbf{k}_{\rm CM}|\cos\theta\sqrt{\gamma^{2}-1}\right)
}
+e^{-\frac{1}{T} \left(E_{\psi_2}\gamma-|\mathbf{k}_{\rm CM}|\cos\theta\sqrt{\gamma^{2}-1}\right)}
\right\}\,,\nonumber
\eea
with $E_{\psi_i}=\sqrt{\mathbf{k}_{\rm CM}^2+m_{\psi_i}^2}$.
In the limit of $m_{\psi_1}=m_{\psi_2}$, eq.~(\ref{eq:sigfinal_full}) coincides with 
eq.~(\ref{eq:sigfinal}) provided in the main text.

If the spin-averaged amplitude itself has no angular dependence, 
$\left|\overline{\mathcal{M}}\right|^2=\left|\overline{\mathcal{M}}\right|^2(s)$, the angular integral 
in eq.~(\ref{eq:sigfinal_full}) can be performed analytically and the full expression for the annihilation
cross section factorises into the standard expression for the cross section in vacuum, 
$\sigma_{\chi\chi\to\psi_1\psi_2}^{\rm CMS}$, and a correction factor $\overline G$:
\be
\label{eq:sig_simp_s}
\sigma_{\chi\chi\to\psi_1\psi_2} (s,\gamma)=
\overline{G}_{\psi_1\psi_2}(\gamma,s)\times\sigma_{\chi\chi\to\psi_1\psi_2}^{\rm CMS}(s)\,.
\ee
Here, 
\bea
\label{eq:gbar_def}
\overline{G}_{\psi_1\psi_2}(\gamma,s)&\equiv& \int_{-1}^{1}\frac{d\cos\theta}{2}G_{\psi_1\psi_2}(\gamma, s, \cos\theta)\\
&=&\frac{1}{(2-A)\log C}\Bigg\{
\tan^{-1}\left(\frac{A-2BC}{2-A}\right)-\tan^{-1}\left(\frac{A-2B}{2-A}\right)\nonumber\\
&&\phantom{\frac{1}{(2-A)\log C}\Bigg\{} 
\tan^{-1}\left(\frac{A-2DC}{2-A}\right)-\tan^{-1}\left(\frac{A-2D}{2-A}\right)
\Bigg\}\,, \label{eq:Gbarpar}
\eea
where 
\be
A\equiv1+\varepsilon^2e^{-\sqrt{s}\gamma/T}\,,\quad
B\equiv\varepsilon e^{-E_{\psi_1}\gamma/T}\,,\quad
C\equiv e^{-|\mathbf{k}_{\rm CM}|\sqrt{\gamma^2-1}/T}\,,\quad
D\equiv\varepsilon e^{-E_{\psi_2}\gamma/T}\,,
\ee
and $\varepsilon\equiv\varepsilon_{\psi_1}=\varepsilon_{\psi_2}$.
The simplified form of the cross section as given in eq.~(\ref{eq:sig_simp_s}) applies in 
particular to all (spin-averaged) annihilations that proceed exclusively via 
$s$-channel processes. It is also directly applicable to the decay of a particle $A$, for which one 
can simply replace all cross sections $\sigma$ in eq.~(\ref{eq:sig_simp_s}) with decay rates 
$\Gamma_A$ (using $s\to m_A^2$ as argument of $\overline{G}$); we used this observation 
in arriving at eq.~(\ref{eq:C_ann_res}) in the main text.

In \ds\ we have implemented a general utility function \code{dsanGbar} that is accessible by
all particle modules, and that returns the quantity $\overline{G}_{\psi_1\psi_2}$
defined in eq.~(\ref{eq:gbar_def}). 
Notably, our implementation relies on a parameterization that is less compact than the one given in 
eq.~(\ref{eq:Gbarpar}), in order to avoid numerical inaccuracies due to significant cancellations 
that can appear between the four terms in the parentheses.

\end{appendix}

\bibliographystyle{JHEP_improved}
\bibliography{biblio.bib}

\end{document}